\documentclass[a4paper,11pt]{article}
\usepackage{amsmath, amssymb, amsfonts,indentfirst,graphicx,color,anysize}
\usepackage{hyperref}
\marginsize{3cm}{3cm}{2cm}{2cm}
\usepackage{float}

\title{
	\includegraphics[width=0.35\textwidth]{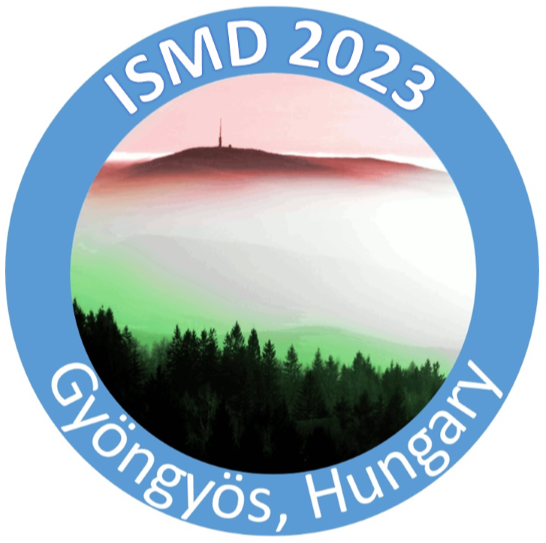}\\[1cm]
	\textbf{Simple L\'evy $\alpha$-Stable Model Analysis of Elastic $pp$ and $p\bar p$ Low-$|t|$ Data from SPS to LHC Energies}}
\author{{T. Cs\"org\H{o}$^{1,2}$, S. Hegyi$^{2}$, and I. Szanyi$^{1,2,3}$}\\[1ex]
	$^1$MATE Institute of Technology,  K\'aroly R\'obert Campus,\\ M\'atrai \'ut 36,
H-3200 Gy\"ongy\"os,  Hungary; 	tcsorgo@cern.ch\\
	$^2$Wigner Research Center for Physics,\\
 P.O. Box 49,
 H-1525 Budapest, Hungary; hegyi.physics@gmail.com\\
        $^3$E\"otv\"os University, Department of Atomic Physics, \\
 P\'azm\'any P. s. 1/A,  H-1117 Budapest, Hungary; iszanyi@cern.ch\\
}
\begin{document}

\maketitle

\vspace{-1.0cm}

\begin{abstract} 

A simple L\'evy $\alpha$-stable (SL) model is used to describe the data on elastic $pp$ and $p\bar p$ scattering at low-$|t|$ from SPS energies up to LHC energies. The SL model is demonstrated to describe the data with a strong non-exponential feature in a statistically acceptable manner. The energy dependence of the parameters of the model is determined and analyzed. The L\'evy $\alpha$ parameter of the model has an energy-independent value of 1.959 $\pm$ 0.002 following from the strong non-exponential behavior of the data.  We strengthen the conclusion that the discrepancy between TOTEM and ATLAS elastic $pp$ differential cross section measurements {arises} only in the normalization and not in the shape of the distribution of the data as a function of $t$. 
We find that the slope parameter has different values for $pp$ and $p\bar p$ elastic scattering at LHC energies. This may be the effect of the odderon exchange or the jump in the energy dependence of the slope parameter in the energy interval \mbox{3 GeV $\lesssim \sqrt s \lesssim$ 4 GeV}.

\end{abstract}

\section{Introduction}  
{
The physics of elastic proton--proton ($pp$) and proton--antiproton ($p\bar p$) scattering can be studied by measuring the differential cross section $d\sigma_{\rm el}/dt$ at a given center of mass (cm) energy $\sqrt{s}$ as a function of the squared four-momentum transfer $t$. The characteristic structure of the $t$-distribution of elastic $pp$ scattering was revealed in the 1970s by experiments performed with the ISR accelerator at CERN \cite{Barbiellini:1972ua,Nagy:1978iw} in the energy range $23~{\rm GeV}\lesssim\sqrt s\lesssim63~{\rm GeV}$. It was found that after the Coulomb-nuclear interference region at very small $|t|$ values, the $d\sigma_{\rm el}/dt$ is nearly exponentially decreasing in the range of $0.01~{\rm GeV}^2\lesssim|t|\lesssim1~{\rm GeV}^2$ and has a characteristic diffractive minimum--maximum (dip-bump) structure in the domain $1~{\rm GeV}^2\lesssim|t|\lesssim2.5~{\rm GeV}^2$.  Beyond the bump, $|t|\gtrsim 3$ $\rm GeV^2$,  the $d\sigma_{\rm el}/dt$ was found to be decreasing according to a power law $|t|^{-n}$ with $n\approx 8$ \cite{Donnachie:1996rq}. Measurements on $pp$ elastic scattering at LHC by TOTEM and ATLAS Collaborations at $\sqrt{s}=$ 2.76, 7, 8, and 13 TeV~\cite{TOTEM:2018psk,TOTEM:2013lle,TOTEM:2015oop,TOTEM:2016lxj,TOTEM:2021imi,TOTEM:2017sdy,TOTEM:2018hki,ATLAS:2014vxr,ATLAS:2016ikn,ATLAS:2022mgx} confirm the structure of the $t$ distribution of elastic $pp$ scattering as observed at ISR and allow for more detailed studies. As the energy rises, the dip-bump structure goes to lower $|t|$ values: at LHC energies, it appears in the range of $0.3~{\rm GeV}^2\lesssim|t|\lesssim1~{\rm GeV}^2$. Remarkably, in elastic $p\bar p$ scattering, only a shoulder-like structure is observed, and no dip is seen \cite{Breakstone:1985pe, UA4:1985oqn,UA4:1986cgb,D0:2012erd}. Otherwise, the $t$-distribution of elastic $p\bar p$ scattering is very similar to that of elastic $pp$ scattering.

In this work, we analyze the low-$|t|$ nearly exponential $d\sigma_{\rm el}/dt$. It was first observed at ISR \cite{Barbiellini:1972ua} and later confirmed at LHC by TOTEM \cite{TOTEM:2015oop,TOTEM:2017sdy,TOTEM:2018hki} that the $pp$ $d\sigma_{\rm el}/dt$ at low-$|t|$ values does not show a purely exponential $Ae^{-B_0|t|}$ structure: there is a change in the slope $B_0$ at around $|t|=0.1$ GeV$^2$. At $\sqrt{s} = 8$ TeV, TOTEM excluded a purely exponential $pp$ $d\sigma_{\rm el}/dt$ in the range of $0.027~{\rm GeV}^2\lesssim|t|\lesssim0.2~{\rm GeV}^2$ with a significance greater than $7\sigma$~\cite{TOTEM:2015oop}. In $p\bar p$ scattering, a change in the slope was observed at SPS at $\sqrt{s} = 540$ GeV and 546 GeV around \mbox{$|t|=0.15$ GeV$^2$} \cite{UA4:1983mlb,UA4:1984skz}.

In the framework of the Regge approach, the non-exponential behavior of the elastic differential cross section was related to the $4m_\pi^2$ branch point of $t$-channel scattering amplitude and, hence, is explained as the manifestation of $t$-channel unitarity  \cite{Cohen-Tannoudji:1972gqd,Anselm:1972ir, Tan:1974gd,Khoze:2000wk,Jenkovszky:2014yea,Fagundes:2015vva, Jenkovszky:2017efs}. According to the findings of Refs.~\cite{Kohara:2019qoq, Kohara:2018wng}, the low-$|t|$ non-exponential behavior of the elastic $pp$ differential cross section can be a consequence of an interplay between the real parts of the Coulomb and nuclear amplitudes.

In order to describe the low-$|t|$ $pp$ $d\sigma_{\rm el}/dt$, the TOTEM Collaboration used the parametrization %MDPI: 1. Please check if all variables need to format consistently (italic/superscript/subscript/etc.) throughout the text. If so, please revise.2. Please ensure that there are no duplicate equations, and all equations appear in numerical order.
 \cite{TOTEM:2015oop}:
\begin{equation}\label{eq:expnonlin}
   \frac{d\sigma}{dt}(s,-t)=a(s)e^{-b_1(s)t+b_2(s)t^2},
\end{equation}
where $a$, $b_1$, and $b_2$ are free parameters to be determined at a given cm energy. In this study, we analyze the low-$|t|$ $pp$ and $p\bar p$ $d\sigma_{\rm el}/dt$ in the energy range $546~{\rm GeV}<\sqrt s<13~{\rm TeV}$ using a simple L\'evy $\alpha$-stable (SL) model, as introduced in \cite{Csorgo:2023pdn}. In the SL model, the low-$|t|$ elastic differential cross section has the form: 
 \begin{equation}\label{eq:SL}
\frac{{\rm d}\sigma}{{\rm d} t}(s,-t) = a_L(s)e^{-|t b_L(s)|^{\alpha_L(s)/2}},
\end{equation}
where the Levy index of stability $\alpha_L(s)$, the optical point parameter $a_L(s)$, and the slope parameter $b_L(s)$ are fit parameters to be determined at a given cm energy.

L\'evy distributions were introduced to high-energy physics in several papers. The common theme of these papers is the application of generalized central limit theorems for the convolution of elementary processes that may have infinite first or second moments (a trivial example of this is the Lorentzian distribution). Due to this property of the elementary processes, the conditions of validity for the classical central limit theorems that lead to Gaussian limiting distributions are not satisfied. % please check to make sure intended meaning has been retained
However, in certain cases, generalized central limit theorems are still valid when the addition of one more elementary process does not modify the shape of the limiting distribution and only modifies the parameters of it. The mathematical theory of generalized central limit theorems were worked out in the 1920s by the French mathematician Paul L\'evy, and the resulting limiting distributions are named after him as L\'evy or L\'evy stable distributions. 

The book of Uchaikin and Zolotarev~\cite{uchaikin2011chance} lists several examples of their applications 
in probabilistic models, related to anomalous diffusion,  astrophysics, biology, chaos, correlated systems and fractals,  financial applications, geology,
physics, radiophysics, and stochastic algorithms, among others. Stable distributions provide solutions
to certain ordinary and fractional differential equations, and the extremely broad range  of their applications 
indicate that L\'evy stable distributions are ubiquitous in nature, as noted and explained by Tsallis and collaborators in Ref.~\cite{Tsallis:1995zz,Prato:1999jj}.
A recent book of J. P. Nolan discussed and also standardized the notation for the theory, numerical algorithms, and statistical methods associated with stable distributions using an accessible, non-technical approach and also highlights many practical applications of L\'evy stable distributions, including their applications
in finance, statistics, engineering, physics, and, in particular, high-energy physics~\cite{nolan2020univariate}.

L\'evy stable distributions were utilized in high-energy physics to explain the applicability of Tsallis distributions~\cite{Tsallis:1995zz} and, in general, power-law tails in the transverse momentum spectra of particles by Wilk and Wlodarczyk in Ref.~\cite{Wilk:1999dr}. They were applied to describe an intermittent behavior in the quark-gluon plasma--hadron gas phase transition % please check to make sure intended meaning has been retained
in Ref.~\cite{Brax:1990jv} and to describe anisotropic dynamical fluctuations in multiparticle production dynamics by
Zhang, Lianshou, and Fang~\cite{Zhang:1995uf}.

Another wave of applications of L\'evy stable distributions in high-energy physics was opened in Ref.~\cite{Csorgo:2003uv}, where univariate and multivariate Bose--Einstein correlation functions were carefully analyzed, and the structure of the peak of these correlation functions was investigated without using the assumption of analyticity at zero relative momentum. In this case, both symmetric and asymmetric, univariate and multivariate L\'evy stable data analysis becomes possible.

Univariate and asymmetric L\'evy distributions were subsequently found to characterize two-particle Bose--Einstein correlation functions in electron--positron collisions at LEP, first using simulated data~\cite{Csorgo:2008ah} and subsequently L3 measurements~\cite{L3:2011kzb}. This univariate L\'evy analysis has been extended to a three-dimensional L\'evy analysis of Bose--Einstein correlation functions on PHENIX preliminary data in Ref.~\cite{Kurgyis:2020vbz}.
Univariate and symmetric L\'evy distributions were found to describe precisely the experimental data of the PHENIX Collaboration of Au+Au collisions at
$\sqrt{s_{NN}} = 200$ GeV in the 0--30\% centrality class~\cite{PHENIX:2017ino}. 
By now, several experiments in high-energy particle and nuclear physics analyze their data with the help of L\'evy distributions, for example, the ATLAS~\cite{Schegelsky:2018tit} and CMS~\cite{CMS:2023xyd} experiments at the Large Hadron Collider (LHC),  the PHENIX~\cite{Lokos:2018dqq} and STAR~\cite{Kincses:2024sin} experiments at the Relativistic Heavy Ion Collider (RHIC),  and the NA61/SHINE experiment at CERN SPS~\cite{Porfy:2019scf,Porfy:2023yii}.
The application of L\'evy stable source distributions to high-energy heavy ion physics has been recently reviewed by Csan\'ad and Kincses in Ref.~\cite{Csanad:2024hva}.

A model-independent expansion technique was proposed by Nov\'ak and colleagues in Ref.~\cite{Novak:2016cyc} to allow for a test of a possible deviation from the L\'evy shape. So far, no such deviations were found in the field of Bose--Einstein correlations as far as we know. However, important deviations were found in a related femtoscopic area called elastic scattering and diffraction, where a quantum interference between the scattered wave and the unscattered incoming plane wave creates a well-measured diffractive interference pattern. The model-independent L\'evy expansion technique was successfully applied to these data and has been utilized to identify hollowness and odderon effects in elastic proton--proton collisions in Refs.~\cite{Csorgo:2019egs,Csorgo:2018uyp}, respectively. 

In the present study, we apply a simple L\'evy $\alpha$-stable model of Equation~(\ref{eq:SL}) to describe low-$|t|$ elastic $pp$ and $p\bar p$ data. The paper is organized as follows. In Section~\ref{sec:basics}, we recapitulate the basic formulae for describing high-energy elastic scattering. In Section~\ref{sec:em_levy}, we detail the emergence of L\'evy $\alpha$-stable distribution in elastic hadron--hadron scattering that leads to the SL model. In Section~\ref{sec:fits}, we analyze the low-$|t|$ $pp$ and $p\bar p$ $d\sigma_{\rm el}/dt$ in the energy range $546~{\rm GeV}<\sqrt s<13~{\rm TeV}$ using the SL model; we present the fits to the data and determine the energy dependencies of the SL model parameters. The results are discussed in Section~\ref{sec:disc} and summarized in Section~\ref{sec:summ}. Simple models of elastic scattering are shortly reviewed in the Appendix.
}

{
\section{Basic Formalism of High-Energy Elastic Scattering} \label{sec:basics}

The formulas that describe high-energy, small-angle scattering of particles are analogous to those describing Fraunhofer diffraction of light by absorbing and refracting obstacles \cite{glauber1959lectures,glauber1970theory,Barone:2002cv}. The characteristics of the ``obstacle'' at a given cm energy and impact parameter $b$ are specified by the profile function,
\begin{equation}\label{eq:profile}
  \Gamma(s,b) = 1-e^{-\Omega(s,b)},  
\end{equation}
where $\Omega(s,b)$ is called the opacity function, which is, in general, a complex quantity~\cite{Glauber:1984su}.

We can define the impact parameter representation of the elastic scattering amplitude $\tilde t_{\rm el} (s, b)$ as
\begin{equation} \label{eq:impamp}
    \Tilde t_{\rm el} (s,b) = i \Gamma(s,b).
\end{equation}
In %MDPI: Please check if need to add indentation. so as below highlighted. AUTHORS: indentation is not needed.
 the general case, one has an impact parameter vector $\vec b$. However, here, we assume azimuthally symmetric interactions, allowing us to fully describe the scattering process using the absolute value of the impact parameter vector $b=|\vec b|$. 

The high-energy, small-angle scattering amplitude in the momentum representation $T(s,t)$ is given as the Fourier transform of  $\tilde t_{\rm el} (s,b)$:
\begin{equation}\label{eq:relPWtoeik_2}
    T_{\rm el}(s,t) = \int d^2\vec b e^{-i\vec \Delta\cdot\vec b }\Tilde t_{\rm el} (s, b),
\end{equation}
where $\vec \Delta$ is the momentum transfer vector. In the high-energy limit applicable in our case, $\Delta=|\vec \Delta|=\sqrt{-t}$. Taking advantage of the azimuthal symmetry, one can integrate out for the azimuthal angle and rewrite Equation~(\ref{eq:relPWtoeik_2}) as
\begin{equation}\label{eq:relPWtoeik_3}
    T_{\rm el}(s,t) = 2\pi \int dbbJ_0(\Delta b)\Tilde t_{\rm el} (s,b),
    \end{equation}
where $J_0$ is the zeroth order Bessel function of the first kind. By inverting Equation~(\ref{eq:relPWtoeik_2}), we can express the amplitude in the impact parameter representation as the integral transformed of the amplitude in the momentum representation:
\begin{equation}\label{eq:relPWtoeik_3}
    \Tilde t_{\rm el} (s,b) =\frac{1}{(2\pi)^2}\int d^2\vec \Delta e^{i\vec \Delta\cdot\vec b }T_{\rm el}(s,\Delta).
\end{equation}
%Equation~(\ref{eq:relPWtoeik_2}) and Equation~(\ref{eq:relPWtoeik_3}) fix our convention for the normalization of the Fourier transform and its inverse. 

Our basic measurable quantity, the differential cross section, is determined by the absolute value squared of the scattering amplitude in momentum representation: 
\begin{equation}\label{eq:diffxsec}
    \frac{{\rm d}\sigma}{{\rm d} t}(s,t) = \frac{1}{4\pi}\left|T_{\rm el}\left(s,t\right)\right|^2.
\end{equation}

When describing scattering processes, we require the unitarity of the scattering matrix. 
At high energies, this unitarity constraint can be expressed as 
\begin{equation}\label{eq:unitb}
    2 Re\Gamma(s,\vec b) = \big|\Gamma(s,\vec b)\big|^2 +  \tilde\sigma_{in}(s, b), 
\end{equation}
where $\tilde\sigma_{in}(s, b)$ is called the shadow profile or inelastic overlap function. 
Utilizing Equation~(\ref{eq:impamp}), Equation~(\ref{eq:unitb}) can be rewritten as 
\begin{equation}\label{eq:unitb_2}
    2 Im \Tilde t_{\rm el}(s,b) = \big| \Tilde t_{\rm el} (s,b)\big|^2 + \tilde\sigma_{in}(s, b).
\end{equation}
Although elastic scattering corresponds to a genuine quantum interference between the elastically scattered wave and the unscattered incoming wave, so it has no probabilistic interpretation, the quantity that describes the inelastic scattering, $\tilde\sigma_{in}(s, b)$, has a probabilistic interpretation, and it can be interpreted as the probability of inelastic scattering at a given energy and impact parameter, as detailed, e.g., in Ref.~\cite{Csorgo:2020wmw}.
%Equations (\ref{eq:unitb},\ref{eq:unitb_2}) are also referred to as generalized optical theorems. When integrating these equations over the impact parameter $b$, the right hand sides yield the elastic and the inelastic cross-sections, that sum up to the total cross-section, while the left hand side corresponds to the imaginary part of the elastic scattering amplitude at the optical point, $t = 0$. This yields the form of the optical theorem given above in Equation~(\ref{eq:optical-theorem}).

\section{Emergence of L\'evy \boldmath{$\alpha$}-Stable Distribution in Elastic Scattering}\label{sec:em_levy}

Realistic models of elastic $pp$ (or $p\bar p$) scattering try to deal simultaneously with the large and small $|t|$ behavior of elastic scattering. One of the key qualitative features of the experimental data is the existence of a unique diffractive minimum and maximum in elastic proton--proton collisions at the TeV energy scale. Glauber's multiple diffractive theory, as implemented in the Bialas--Bzdak (BB) model \cite{Bialas:2006qf}, relates the number of diffractive minima to the basic structure of the proton in elastic scattering. Namely, if the proton behaves as a weakly bound quark--diquark state, denoted as $p = (q,d)$, it corresponds to one-hole on one-hole scattering and has only a unique diffractive minimum in the experimentally available four-momentum transfer range. On the other hand, if the diquark can be decomposed as a weakly bound quark--quark state, this leads to the $p= (q, (q,q)) $ structure of the proton, and in this case, several diffractive minima are predicted.  The BB model thus comes in two variants: the proton can be modeled either as a weakly bound quark--diquark state, $p = (q,d)$, or the diquark can be resolved as a weakly bound quark--diquark state, $p = (q, (q,q))$.

The original $p= (q,d)$ and $p = (q, (q, q))$ BB models~\cite{Bialas:2006qf} were formulated by neglecting the real part of the elastic scattering amplitude. The real part was added in a unitary manner in Refs.~\cite{Nemes:2015iia,Csorgo:2020wmw}, leading to the so-called Real-extended Bialas--Bzdak model (ReBB). In the quark--diquark model of elastic $pp$ scattering, the $p = (q,d)$ variant has only a single diffractive minimum, and this variation of the ReBB model gives a statistically acceptable description to the proton--proton ($pp$) and proton--antiproton ($p\bar p$) elastic scattering data in a limited kinematic range \cite{Csorgo:2020wmw, Szanyi:2022ezh} that includes the diffractive interference (minimum and maximum) region but does not include the low-$|t|$ domain, where a strong non-exponential shape characterizes the experimental data. % in the TeV energy range \cite{TOTEM:2015oop,Csorgo:2016qyr}. % please check to make sure intended meaning has been retained
Although the ReBB model also featured a non-exponential behavior at low $-t$ values, this was obtained as a small violation of a nearly exponential behavior of elastic $pp$ scattering that was not strong enough to describe the experimental observations~\cite{Nemes:2015iia,Csorgo:2020wmw}. This can be attributed to the assumption of BB models that all relative distance scales are Gaussian random variables. % in the BB model. 
As detailed below, if we assume Gaussian distributions and apply a central limit theorem to the tail of the inelastic collision $b$-distribution, $\sigma_{\rm in}(s,b)$, a Gaussian model is obtained with an exponential behavior in the diffractive cone region.

%This feature (ie the dip-bump) is difficult to describe in simple models like grey Gaussian (that has no diffractive interference structure) or the black and grey disc models, that have infinitely many such interference patterns.

%We can write $\Tilde t_{\rm el}(s,b)$ in terms of $\tilde\sigma_{in}(s,\vec b)$ as a solution of the unitarity equation, Equation~(\ref{eq:unitb_2}). 

For elastic scattering at high energies, we know from experimental data that the elastic scattering amplitude is dominantly imaginary at small values of $|t|$; see, for example, Ref.~\cite{TOTEM:2017sdy} for a recent summary of measurements of $\rho(s)$, the ratio of the real to imaginary part of the elastic scattering amplitude at a vanishing four-momentum transfer. In this case, as it follows from Equation~(\ref{eq:impamp}), the profile function $\Gamma(s,\vec b)$ is dominantly real. Neglecting its imaginary part, one can write $\Gamma(s,\vec b)$ as a particular solution of Equation~(\ref{eq:unitb}) in the form 
\begin{eqnarray}\label{eq:gamma_1}
    \Gamma(s,\vec b)=1 - \sqrt{1-\tilde\sigma_{in}(s, b)}.
\end{eqnarray}
At %MDPI: Please check if need to add indentation. AUTHORS: indentation is not needed.
 large $b$, both $\tilde\sigma_{in}(s,\vec b)$ as well as  $\Gamma(s,\vec b)$, are expected to be small. This selects the negative sign in the right-hand side of 
the above equation.
If the elastic scattering is described in the impact parameter space by the small but analytic function of $\tilde\sigma_{in}(s, b) \ll 1$,
a large $b$ approximation of Equation~(\ref{eq:gamma_1}) is given as
\begin{eqnarray}\label{eq:gamma_2} 
    \Gamma(s,\vec b)\underset{{\rm large}~b}{\simeq}\frac{1}{2}\tilde\sigma_{in}(s, b) .
\end{eqnarray}

%The imaginary part of the elastic scattering amplitude is the dominant part, whereas the real part can be considered as a smaller correction. Neglecting the real part of the amplitude, the amplitude is fully imaginary and we can write
%\begin{equation}\label{eq:im_amplitude}
%    \tilde t_{el}(s,\vec b) = i\left(1-\sqrt{1-\tilde\sigma_{in}(s,\vec b)}\right).
%\end{equation}

%However, in a model where the amplitude does not have a real part, the characteristic minimum--maximum region of the $pp$ differential cross section can not be described properly. In Ref.~\cite{Nemes:2015iia}, the elastic scattering amplitude was extended with a real part in a way that the unitarity constraint is fulfilled. This amplitude reads as:
%\begin{equation}\label{eq:uBB_ansatz_sub}
%t_{el}(s,\vec b)=i\left(1-e^{i\, \alpha\, \tilde\sigma_{in}(s,\vec b)}\sqrt{1-\tilde\sigma_{in}(s,\vec b)}\right),
%\end{equation}
%where $\alpha$ is a free parameter to be fitted to the data. In the case of $\alpha = 0$, 
%Equation~(\ref{eq:uBB_ansatz_sub}) reduces to Equation~(\ref{eq:im_amplitude}), i.e., to a scattering amplitude that has a vanishing real part.

%This ReBB model worked well around the diffractive interference (minimum and maximum) region, it correctly produced only one diffractive minimum and one diffractive maximum in its $p= (q,d)$ version, however, the model was not working well in the diffractive cone (low $-t$) region. The $\sqrt{s} = 8$ TeV data of the TOTEM experiment in the low $-t$ region indicated a significantly non-exponential structure of small $-t$ elastic scattering ~\cite{TOTEM:2015oop}. 

In this work, we limit our studies to elastic $pp$ scattering at small $-t$, where a nearly exponential behavior, the so-called diffractive cone phenomena, is observed. This can be obtained from several different approaches (see the Appendix for a short review of simple models of elastic scattering). Let us now consider a derivation that is based on the validity of the central limit theorem of probability distributions.

In a nearly exponential $-t $ region, it is customary to use a Gaussian approximation for the shadow profile function, $\tilde\sigma_{\rm in}$.
Such a behavior can be obtained from the fact that inelastic scattering may have a probabilistic interpretation, and if these scatterings are obtained as 
convolutions of elementary scattering processes, where each of the convoluted elementary distributions has a finite mean and a finite variance,
then, in the limit of a large number of convolutions, the resulting net distribution for inelastic scattering has an asymptotically Gaussian form, according to the central limit theorem of probability distributions:
\begin{eqnarray}\label{eq:profgauss}
\tilde\sigma_{in}(s, b) =2\Gamma(s,b) =  % \frac{2 c_G(s)}{(2\pi)^2}\int d^2 \Delta e^{i{\vec \Delta}\cdot \vec b}e^{-\frac{1}{2}\Delta^2 R^2_G(s)}=
%\sqrt{{4\pi}{A(s)}}
%\frac{G_r(s)}{2 %\pi R^2_G(s)}
2 c_G(s) %\pi R^2_G(s)
\exp\left({-\frac{b^2}{2 b_G(s)}}\right),
\end{eqnarray}
where % $A(s)$ is an energy-dependent overall normalization factor, which interpretation is detailed below, 
$c_G(s)$ is an $s$-dependent factor, and $b_{G}(s)$ is an $s$-dependent slope parameter (for more details, see the Appendix).

In the small $-t$ approximation, corresponding to large values of the impact parameter  $b$, the Gaussian model yields the following elastic scattering amplitudes:
\begin{eqnarray}
     \Tilde t_{\rm el} (s,b) & = &  i c_G(s) %\sqrt{{4\pi}{A(s)}}
                       %\frac{1}{2}
                        \exp{\left(-\frac{b^2}{2 b_G(s)}\right)} , \\
      T_{\rm el}(s,t)  & = & % \int d^2\vec b e^{-i\vec \Delta\cdot\vec b }\Tilde t_{\rm el} (s,\vec b) \, = 
               2 \pi i c_G(s)  b_G(s)
               \exp{  \left(\frac{t b_G(s)}{2}\right)}.
\end{eqnarray}

Given that small-$|t|$ scattering corresponds to large $b$ scattering, the Gaussian approximation of Equation~(\ref{eq:profgauss}) corresponds to the assumption of the validity of  a central limit theorem for scattering at large impact parameters: if many small random scatterings with finite means and root mean squares are convoluted to describe the inelastic scattering at large impact parameters, a Gaussian source emerges due to the validity of the central limit theorem of probability distributions, 
without reference to the particular details of the elementary inelastic scattering processes. 

This so-called gray Gaussian model, together  with Equations~(\ref{eq:impamp}), (\ref{eq:relPWtoeik_2}), (\ref{eq:diffxsec}), and (\ref{eq:gamma_2}), leads to an exponential small-$|t|$ differential cross section 
\begin{equation}
\label{eq:SL0}
\frac{{\rm d}\sigma}{{\rm d} t}(s,t)  = 
a_G(s) \,e^{t b_G(s)},
\end{equation}
where the fit parameter $a_G(s)=\pi c_G^2(s) b^2_G(s)$ is the optical point parameter, as this is the value of the differential cross section at the optical point, corresponding to an extrapolation to $t=0$. 

The above-given Gaussian model has two shortcomings. One of them is obvious: given that the experimental data on elastic  $pp$ scattering have a single diffractive minimum and
the Gaussian model has no diffractive minimum, it is clear that the domain of validity of this model does not extend to the vicinity of the diffractive minimum.
Furthermore,  a nearly exponential cone is observed experimentally at low values of the squared four-momentum transfer $-t$. 
This leads to the second, more subtle shortcoming of this Gaussian source model:
as long as the experimental data are not precise enough to see deviations from an exponentially falling diffractive cone, the Gaussian model remains adequate. However, as we  have already entered the domain of high statistic and precise elastic $pp$ scattering at LHC energies of 8 and 13 TeV, a subtle  but statistically significant
deviation from an exponential behavior has been discovered by the TOTEM Collaboration \cite{TOTEM:2015oop,TOTEM:2017sdy,TOTEM:2018hki}. % at 8  TeV LHC energy~\cite{TOTEM:2015oop,Csorgo:2016qyr}. 
This experimental discovery rules out, with a statistical significance much greater than 5$\sigma$, models that rely on the existence of an exactly exponential diffractive cone in elastic $pp$ scattering. Thus, models with a nearly Gaussian tail in their probability distribution of inelastic $pp$ scattering are ruled out by TOTEM data at $\sqrt{s} = 8$ TeV.

In the ReBB model, the assumed quark and diquark constituents of the proton have Gaussian parton distributions, and also, the distance between these constituents has a Gaussian shape. Consequently, in the ReBB model, $\sigma_{in}$ is a sum of convolutions of Gaussian-shaped terms; hence, this model leads to a nearly exponential differential cross section in the diffractive cone: to the leading order, the shape is exponential, which is modified by weakly non-exponential correction terms, as detailed in Refs.~\cite{Csorgo:2020wmw,Nemes:2015iia}.

In a recent work \cite{Csorgo:2023pdn}, we have formulated Glauber's multiple diffractive theory for a case where the elementary distributions in proton--proton collisions have a power-law type, L\'evy tail. In particular, we have formulated the real extended L\'evy $\alpha$-stable generalized Bialas--Bzdak (LBB) model as the generalization of the ReBB model~\cite{Nemes:2015iia,Csorgo:2020wmw} of elastic $pp$ and $p\bar p$ scattering.  
However, to apply the full LBB model to analyze the data, one needs to solve the problem of integrating products of two-dimensional L\'evy $\alpha$-stable distributions, and access to relatively high computing resources is necessary. As a temporal solution, we introduced approximations that are valid at the low-$|t|$ domain of elastic scattering. This led~\cite{Csorgo:2023pdn} to the to the SL model, as given by Equation~(\ref{eq:SL}). In Ref.~\cite{Csorgo:2023pdn}, we have demonstrated that the SL model describes the non-exponential low-$|t|$ differential cross section of $pp$ scattering at 8 TeV in a statistically acceptable manner. In contrast, the original form of the ReBB model with Gaussan-shaped terms in $\tilde\sigma_{in}$ could not reproduce this strong non-exponential feature of the data.

The Gaussian distribution is the $\alpha=2$ special case of the L\'evy $\alpha$-stable distribution. The LBB model with L\'evy $\alpha$-stable parton and distance distributions may reproduce the strong non-exponential behavior seen in the low-$|t|$ data. In the case of the LBB model, the leading terms in $\sigma_{in}$ are a sum of L\'evy $\alpha$-stable shaped terms. Based on generalized central limit theorems, these scatterings correspond to power-law tailed probability distributions, where the second and possibly even the first moment of the distributions is infinite. This leads to a non-analytic, stretched exponential shape of the elastic scattering amplitude at small values of the four-momentum transfer and a strongly non-exponential behavior of elastic proton--proton ($pp $) scattering at the TeV energy scale.

The SL model of Ref.~\cite{Csorgo:2023pdn} given in Equation~(\ref{eq:SL}) is obtained by choosing $\tilde\sigma_{in}$ to have a L\'evy $\alpha$-stable shape,
\begin{equation}\label{eq:sigmain_levy}
\tilde\sigma_{in}(s,\vec b)=2\Gamma(s,b)=  2c_L(s)\int d^2 \vec \Delta e^{i{\vec \Delta} \cdot\vec b}e^{-\frac{1}{2}\left|\Delta^2b_L(s)\right|^{\alpha_L(s)/2}},
\end{equation}
where $c_L(s)$ is an energy-dependent overall normalization factor, $b_L(s)$ is the L\'evy slope parameter, and  $\alpha_L(s)$ is the L\'evy-$\alpha$ parameter. 
Equation~(\ref{eq:sigmain_levy})
with Equation~(\ref{eq:gamma_2}), and Equations~(\ref{eq:impamp}), (\ref{eq:relPWtoeik_2}), and (\ref{eq:diffxsec}) lead to the SL model, i.e., %MDPI: We removed italic of it. Please check and confirm. AUTHORS: OK.
 a non-exponential low-$|t|$ differential cross section of the form given in Equation~(\ref{eq:SL}). The $\alpha_L=2$ {case} corresponds to a Gaussian profile, as given by Equation~(\ref{eq:profgauss}), and an exponential differential cross section, as given by Equation~(\ref{eq:SL0}). {In the case of $0 < \alpha_L < 2$, the impact parameter profiles, $\tilde\sigma_{\rm in}$ and $\Gamma$, are L\'evy $\alpha$-stable distributed at high-$b$, having a long tail, and the differential cross section is non-exponential as at low-$|t|$.}

}

\section{SL Model Analysis of Elastic \boldmath{$pp$} and \boldmath{$p\bar p$} Low-\boldmath{$|t|$} Data} \label{sec:fits}

{
We present in this section the results of the SL model fits to the low-$|t|$ $pp$ and $p\bar p$ $d\sigma_{\rm el}/dt$ in the energy range $546~{\rm GeV}<\sqrt s<13~{\rm TeV}$. First, we detail the fit method, then we show the fit results and determine the energy dependencies of the fit parameters of the SL model.

We performed the fitting procedure} by using a $\chi^{2}$ definition, which relies on a method developed by the PHENIX Collaboration \cite{Adare:2008cg}. This $\chi^2$ definition is equivalent to the diagonalization of the covariance matrix of statistical and systematic uncertainties if the experimental errors are separated into three different types: 
\begin{itemize}
    \item type \textit{a}: point-to-point varying uncorrelated systematic and statistical errors;
    \item type \textit{b}: point-to-point varying and 100\% correlated systematic errors;
     \item type \textit{c}: point-independent, overall correlated systematic uncertainties that scale all the data points up and down by the same factor.
\end{itemize}

We categorized the available experimental uncertainties into these three types as follows: horizontal and vertical $t$-dependent statistical errors (type \textit{a}), horizontal and vertical $t$-dependent systematic errors (type \textit{b}), and overall normalization uncertainties (type \textit{c}). 
The $\chi^{2}$ function used in the fitting procedure is:
\begin{eqnarray}
 \chi ^{2}&=&\left(\sum _{i=1}^{N}\frac{ \left( d_{i}+ \epsilon _{b} \widetilde\sigma _{bi}+ \epsilon _{c} \sigma _{c}d_{i}-m_{i} \right) ^{2}}{\widetilde{ \sigma }_{i}^{2}}\right)+ \epsilon _{b}^{2}+ \epsilon _{c}^{2},
 \label{eq:chi2-final}
\end{eqnarray}
where 
 \begin{equation}
    \widetilde{ \sigma }_{i}^{2}= \widetilde\sigma _{ai} \left( \frac{d_{i}+ \epsilon _{b} \widetilde\sigma _{bi}+ \epsilon _{c}\sigma _{c}d_{i}}{d_{i}} \right),
\end{equation}
\begin{equation}
     \widetilde\sigma_{ki} =\sqrt{\sigma _{ki}^2+ (d^{\prime}_{i} \delta_{k}t_{i})^2}, \ \ k\in\{a,b\}, \ \  d^{\prime}(t_{i})=\frac{d_{i+1}-d_{i}}{t_{i+1}-t_{i}},
     \end{equation}
$N$ is the number of fitted data points, $d_{i}$ is the $i$th measured data point, and $m_{i}$ is the corresponding value calculated from the model; $ \sigma_{ki}$ is the type $k\in\{a,b\}$ error of the data point $i$, $\sigma_c$ is the type \textit{c} overall error given in percents, $d^{\prime}_{ij}$ denotes the numerical derivative in point $t_{i}$ with errors of type $k\in\{a,b\}$, denoted as $\delta_{k}t_{i}$; $\epsilon_l$ is the correlation coefficient for a type $l\in\{b,c\}$ error. These correlation coefficients are fitted to the data and must be considered as both free parameters and data points not altering the number of degrees of freedom. The $\chi^2$ definition, Equation~(\ref{eq:chi2-final}), was utilized and further detailed in Ref.~\cite{Csorgo:2020wmw}.

The SL model was fitted using the above detailed $\chi^2$ definition, Equation~(\ref{eq:chi2-final}), to all the available $pp$ and $p\bar{p}$ differential cross section data in the kinematic range of 0.54~TeV $\leq\sqrt{s}\leq 13$ TeV and 0.02 GeV$^2$ $\leq -t\leq0.15$ GeV$^2$. {In total, eleven $pp$ and $p\bar p$ datasets were included in the analysis.} The values of the parameters of the model at different energies as well as the confidence levels of the fits and the data sources are shown in Table~\ref{tab:1}. { We regard a fit by a model to be a statistically acceptable description in the case of \mbox{0.1\% $\leq$ CL $\leq$ 99.9\%}. One can see that the confidence level (CL) values range from 8.8\% to 96\%, implying that the SL model represents the data in a statistically acceptable manner.}

\begin{table}[!hbt]
    \centering
    \begin{tabular}{cccccc}
   \hline\hline\noalign{\smallskip}
        $\sqrt{s}$ [GeV] &data from& $\alpha_L$ & $a$ [mb/GeV$^2$] & $b$ [GeV$^{-2}$] & CL (\%)   \\ \noalign{\smallskip}\hline \hline\noalign{\smallskip}
546& UA4 \cite{UA4:1984skz}&1.93 $\pm$ 0.09&209 $\pm$ 15&15.8 $\pm$ 0.9&18.1 \\
1800& E-710 \cite{E-710:1990vqb}&2.0 $\pm$ 1.5&270 $\pm$ 24&16.2 $\pm$ 0.2&77.1\\
2760&TOTEM \cite{TOTEM:2018psk}&1.6 $\pm$ 0.3&637 $\pm$ 25&28 $\pm$ 11&20.5\\
7000 &TOTEM \cite{TOTEM:2013lle}&1.95 $\pm$ 0.01&535 $\pm$ 30&20.5 $\pm$ 0.2&8.8\\
7000 &ATLAS \cite{ATLAS:2014vxr} &1.97 $\pm$ 0.01&463 $\pm$ 13&19.8 $\pm$ 0.2&96.0\\
8000 &TOTEM \cite{TOTEM:2015oop}&1.955 $\pm$ 0.005&566 $\pm$ 31&20.09 $\pm$ 0.08&43.9\\
8000 &TOTEM \cite{TOTEM:2016lxj}&1.90 $\pm$ 0.03&582 $\pm$ 33&20.9 $\pm$ 0.4&19.6\\
8000 &ATLAS \cite{ATLAS:2016ikn}&1.97 $\pm$ 0.01&480 $\pm$ 11&19.9 $\pm$ 0.1&55.8\\
13,000 
 &TOTEM \cite{TOTEM:2017sdy}&1.959 $\pm$ 0.006&677 $\pm$ 36&20.99 $\pm$ 0.08&76.5\\
13,000 &TOTEM \cite{TOTEM:2018hki}&1.958 $\pm$ 0.003&648 $\pm$ 95&21.06 $\pm$ 0.05&89.1\\
13,000 &ATLAS \cite{ATLAS:2022mgx}&1.968 $\pm$ 0.006&569 $\pm$ 17&20.84 $\pm$ 0.07&29.7\\ 
\hline\hline
    \end{tabular}  
\caption{The values of the parameters of the SL model at different energies from half TeV up to 13~TeV in the four-momentum transfer range 0.02 GeV$^2$ $\leq -t\leq0.15$ GeV$^2$. The last column shows the confidence level of the fit to the data at different energies. }\label{tab:1}
\end{table}

Using the values of the parameters of the model at different energies given in Table~\ref{tab:1}, we determined the energy dependence of these parameters.

Table~\ref{tab:1} indicates that the TOTEM datasets at $\sqrt{s} = $ 7,  8 and 13 TeV, as well as the ATLAS dataset at $\sqrt{s} =  13$ TeV feature a strongly non-exponential shape, with  $\alpha_L$ significantly less than 2. The other datasets provide a less precise value for this L\'evy exponent.

\begin{figure}[!hbt]
	\centering
\includegraphics[width=0.88\linewidth]{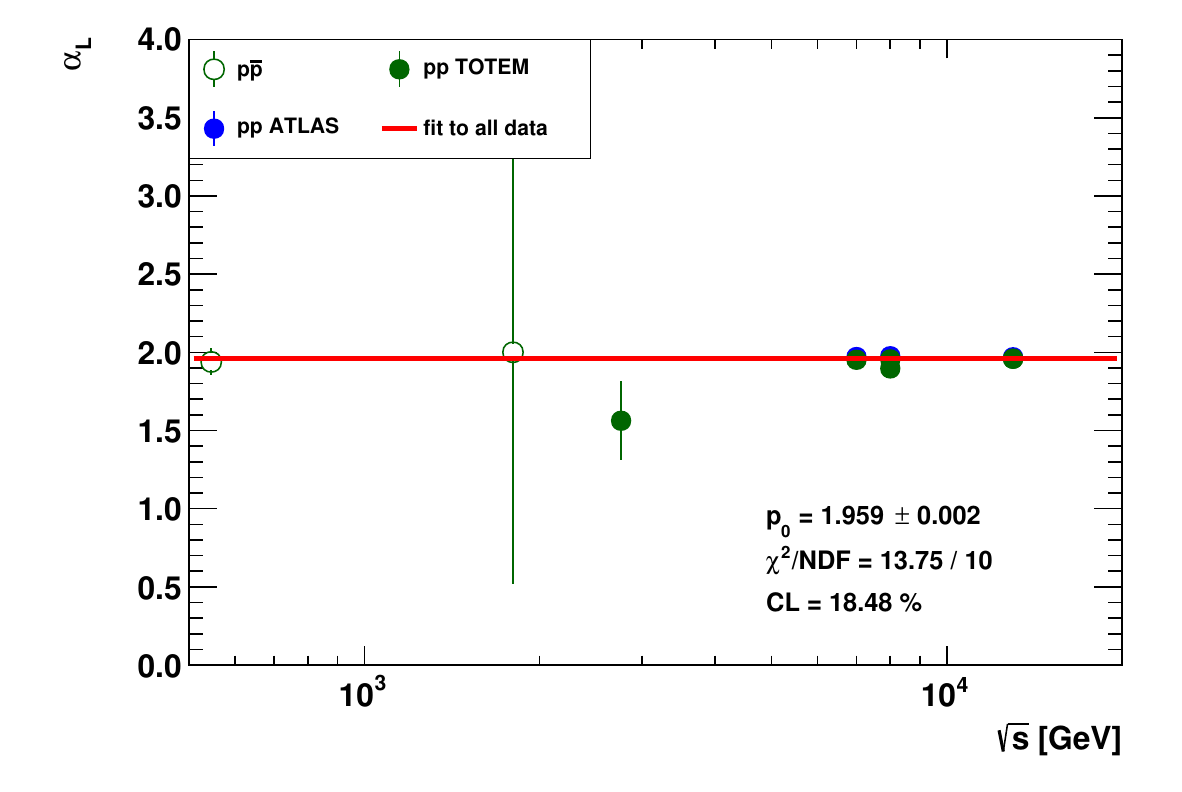}
\vspace{-0.2cm}
	\caption{The %MDPI: we moved the position to reduce the blank page. AUTHORS: OK.
 values of the $\alpha_L$ parameter of the SL model at different energies from half TeV up to 13 TeV. The $\alpha_L$ parameter of the model is energy-independent:  its values at different energies can be fitted with a constant, 1.959 $\pm$ 0.002.}
	\label{fig:1}
\end{figure}

The $\alpha_L(s)$ parameters can be fitted with an energy-independent constant  $\alpha_L$ value, as shown in Figure~\ref{fig:1}. This average, constant value of the $\alpha_L$ parameter is consistent with all the measurements, with  $\alpha_L = 1.959 \pm 0.002$. Although this average value is close to the Gaussian $\alpha_L = 2$ case, {which corresponds} to an exponentially shaped cone of the differential cross section of elastic scattering, its error is small, and thus, the constant value of $\alpha_L$ is significantly less than 2. This indicates that a strongly non-exponential SL model is consistent with all the low-$|t|$ datasets cited in Table~\ref{tab:1}.

\begin{figure}[!hbt]
	\centering
\includegraphics[width=0.9\linewidth]{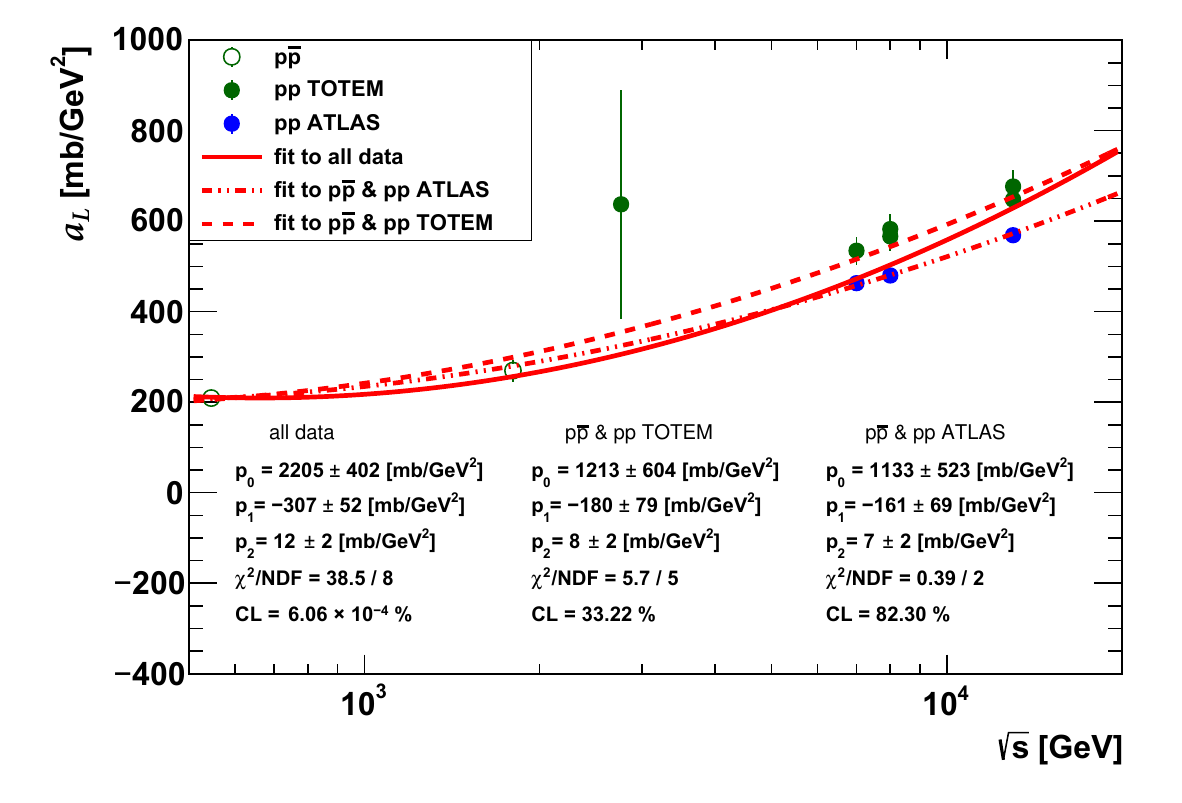}
\vspace{-0.3cm}
	\caption{The %MDPI: 1. Please change the hyphen (-) into a minus sign ($-$, "U+2212"). e.g., "-1" should be "$-$1". 2. Please change the terms into scientific notations in the figure,  e.g., "$8 \times 10^{3}$", not "8E3". AUTHORS: Done.
 values of the optical point parameter of the SL model at different energies from half TeV up to 13 TeV.}
	\label{fig:2}
\end{figure}

{We know that the optical point parameter is proportional to the square of the total cross section \cite{Barone:2002cv,Csorgo:2023pdn}, $a(s)\sim\sigma^2_{\rm tot}(s)$, and the total cross section is related to the square of the size parameter \cite{Csorgo:2023pdn}, $\sigma_{\rm tot}(s)\sim R^2(s)$. This leads to the relation $a(s)\sim R^4(s)$ (see more details also in the Appendix).  

The ``geometric picture'', based on a series of studies \cite{Cheng:1969eh,Cheng:1969bf,Cheng:1969ac,Cheng:1987ga,Bourrely:1978da,Bourrely:2014efa}, motivates the $\ln^2(s/s_0)$ behavior of the total cross section and the $\ln(s/s_0)$ behavior of the size parameter. In addition, the leading $\ln^2(s/s_0)$ term of hadronic total cross sections was obtained from lattice QCD calculations \cite{Giordano:2012mn, Giordano:2013iga}. In the Donnachie--Landshoff approach of hadronic cross sections \cite{donnachie_dosch_landshoff_nachtmann_2002}, the energy dependence is described by a sum of powers of $s$, $\sum_i c_i(s/s_0)^{\delta_i}$. However, it was discussed in Refs.~\cite{Patrignani_2016,Martynov:2007dy,Martynov:2007kn,Cudell:2005sg,Cahn:1982nr,Martynov:2013ana} that models with a $\ln^2(s/s_0)$ asymptotic term work much better than those with $\ln(s/s_0)$ or $(s/s_0)^\delta$ asymptotic terms.

In Ref.~\cite{Csorgo:2020wmw}, the size parameter was successfully parameterized by a $\sum_{i=0}^{1}p_i\ln^i(s/s_0)$ functional form in the energy range of 546 GeV $\leq \sqrt{s}\leq$ 8 TeV. This would suggest a $\sum_{i=0}^{4}p_i\ln^i(s/s_0)$ functional form for the energy dependence of the optical point parameter. However, in our analysis with the SL model, we find that the energy dependence of the optical point $a_L(s)$ for $p\bar p$ and ATLAS or for $p\bar p$ and TOTEM data in the energy range 0.54~TeV $\leq\sqrt{s}\leq 13$ TeV
is compatible with a quadratically logarithmic shape, 
\begin{equation}\label{eq:quad}
    a(s) = p_0 + p_1\ln\left(\frac{s}{1~{\rm GeV^2}}\right) + p_2\ln^2\left(\frac{s}{1~{\rm GeV^2}}\right),
\end{equation}
i.e., the corrections of the $\ln^3(s/s_0)$ and $\ln^4(s/s_0)$ terms are smaller than the current experimental precision. Our result for the energy dependence of the optical point $a_L(s)$ is shown in Figure~\ref{fig:2}.} 

\vspace{-6pt}

For $p\bar p$ and ATLAS data, the values of the parameters in Equation~(\ref{eq:quad}) are $p_0=1213\pm604$ mb/GeV$^2$, $p_1=-180\pm79$ mb/GeV$^2$, and $p_2=8\pm2$ mb/GeV$^2$, resulting in a confidence level of $33.22\%$. For $p\bar p$ and TOTEM data, the parameter values are $p_0=1133\pm523$ mb/GeV$^2$, $p_1=-161\pm69$ mb/GeV$^2$, and $p_2=7\pm2$ mb/GeV$^2$, resulting in a confidence level of $82.30~\%$. A fit by the parametrization Equation~(\ref{eq:quad}) that includes $a$ parameter values for all data---$p\bar p$, ATLAS, and TOTEM---is statistically not acceptable since its confidence level is 6.06$~\times~10 ^{-4}\%$.

\vspace{-3pt}

\begin{figure}[!hbt]
	%\centering
\includegraphics[width=0.99\linewidth]{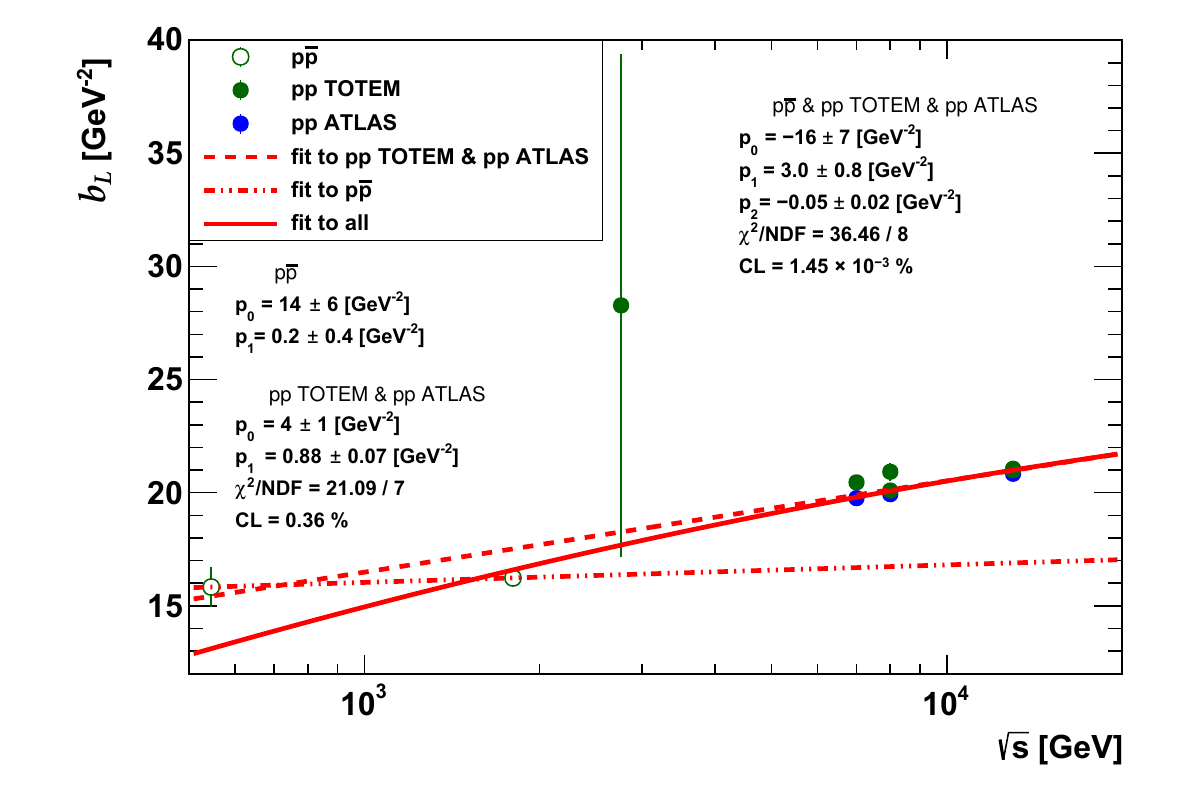}
\vspace{-0.1cm}
	\caption{The %MDPI: Please change the hyphen (-) into a minus sign ($-$, "U+2212"). e.g., "-1" should be "$-$1". AUTHORS: Done.
 values of the slope parameter of the SL model at different energies from half TeV up to 13 TeV.}
	\label{fig:3}
\end{figure}

{As discussed in the Appendix, the slope parameter is related to the size parameter as $b(s)=R^2(s)$. This would suggest a $\sum_{i=0}^{2}=p_i\ln^i(s/s_0)$ functional form for the energy dependence of the optical point parameter. However, we find that for  ATLAS and TOTEM $pp$ data, the energy dependence of the $b_L$ parameter is compatible with a linearly logarithmic shape, 
\begin{equation}\label{eq:lin}
    b(s) = p_0 + p_1\ln\left(\frac{s}{1~{\rm GeV^2}}\right),
\end{equation}
with $p_0=4\pm1$ GeV$^{-2}$ and $p_1=0.88\pm0.07$ GeV$^{-2}$, resulting in a confidence level of $0.36\%$, as illustrated in Figure~\ref{fig:3}. This result, when taken together with the results of Figure~\ref{fig:1} and Figure~\ref{fig:2}, suggests that ATLAS and TOTEM data in the low $-t$ region have a consistent non-exponential shape but differ in their overall normalization. A linearly logarithmic shape for the slope parameter also follows form the one-pomeron exchange Regge model~\cite{Barone:2002cv}.

The values of the $b_L$ parameter for $p\bar p$ data lie on the line given by Equation~(\ref{eq:lin}) with the parameters $p_0=14\pm6$ GeV$^{-2}$ and $p_1=0.2\pm0.4$ GeV$^{-2}$. These values are significantly different from the values of linearity for elastic $pp$ collisions, $p_0=4\pm1$ GeV$^{-2}$ and $p_1=0.88\pm0.07$ GeV$^{-2}$.
The fit for the $b_L$ parameter values of all data---$p\bar p$, ATLAS, and TOTEM---even by the quadratic parametrization Equation~(\ref{eq:quad}) is statistically not acceptable, as it has too small of a confidence level of~$1.45\times 10^{-3}\%$.}

\vspace{-3pt}

\section{Discussion} \label{sec:disc}

In this work, we fitted the low-$|t|$ elastic $pp$ and $p\bar p$ differential cross section in the center of mass energy range of 0.54~TeV $\leq\sqrt{s}\leq 13$ TeV. To do this, we used the SL model as defined by Equation~(\ref{eq:SL}). { Another popular empirical parametrization for the low-$|t|$ non-exponential differential cross section is Equation~(\ref{eq:expnonlin}).} The effect of the quadratic term in the exponent of Equation~(\ref{eq:expnonlin}) is reproduced in our model by an $\alpha_L$ parameter value less than 2. 

An exponential differential cross section corresponds to a Gaussian impact parameter profile. The Gaussian distribution is the $\alpha_L=2$ special case of the more general L\'evy $\alpha$-stable distributions. {The experimentally observed non-exponential differential cross section at low-$|t|$ indicates that the impact parameter profiles, $\tilde\sigma_{in}$ and $\Gamma$, rather have L\'evy $\alpha$-stable shapes at high-$b$, resulting in the SL model given by Equation~(\ref{eq:SL})}. Accordingly, it may be more natural to use Equation~(\ref{eq:SL}) instead of Equation~(\ref{eq:expnonlin}) to model the experimental data. { L\'evy-$\alpha$-stable-shaped source functions are extensively used in the analysis of data on heavy ion collisions, as detailed in the Introduction.}

\begin{figure}[!hbt]
	\centering
\includegraphics[width=0.84\linewidth]{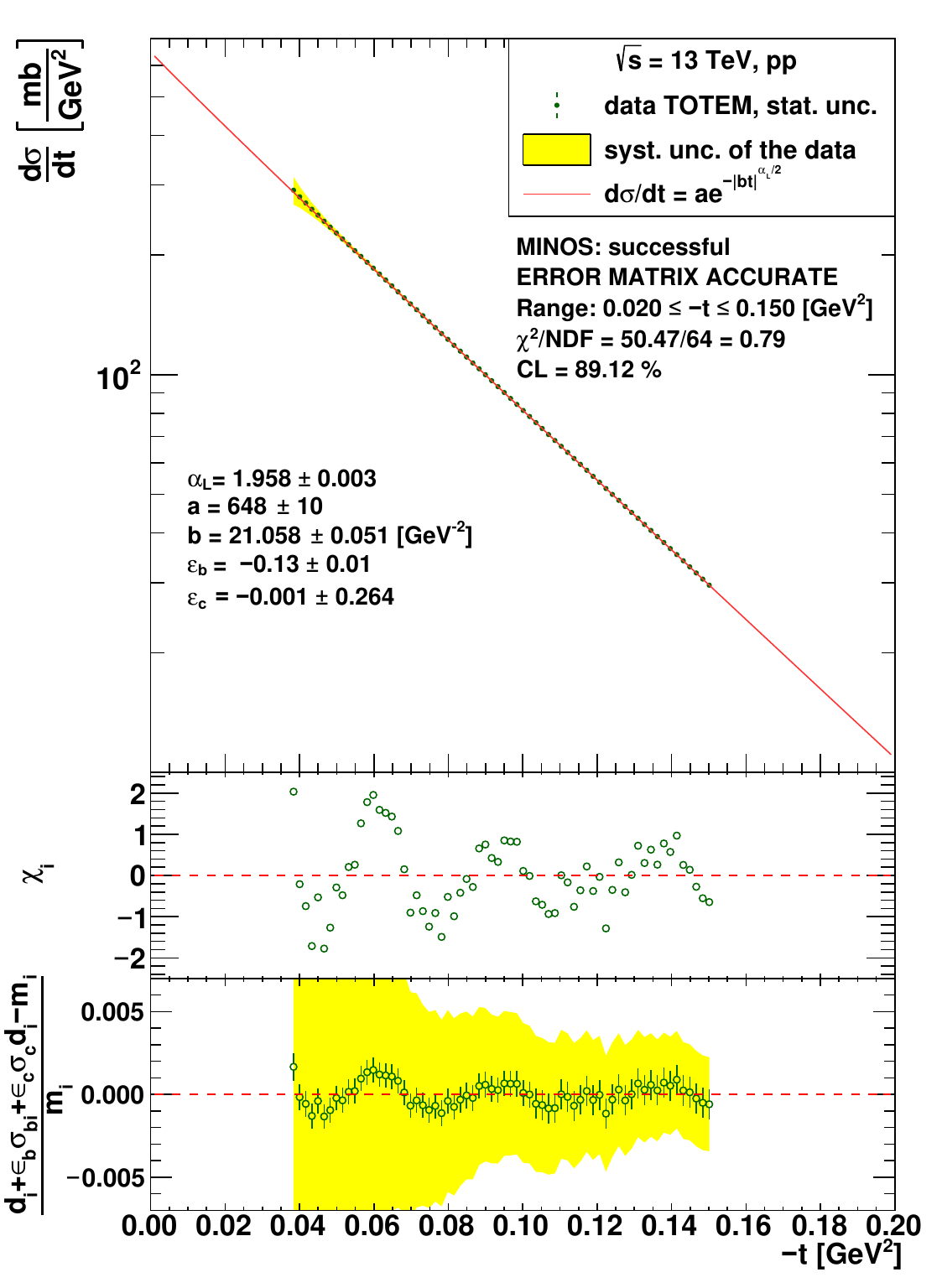}
	\caption{Fit %MDPI: 1. Please change the hyphen (-) into a minus sign ($-$, "U+2212"). e.g., "-1" should be "$-$1". 2. Please change the terms into scientific notations in the figure,  e.g., "$8 \times 10^{3}$", not "8E3". AUTHORS: Done.
 to the low-$|t|$ $pp$ differential cross section data measured by TOTEM  at $\sqrt s = $ 13 TeV~\cite{TOTEM:2018hki}, with the SL model defined by Equation~(\ref{eq:SL}). The differential cross section data with the fitted model curve as well as the values of the fit parameters and the fit statistics are shown in the top panel. {The middle panel shows the $\{\chi_i\}$ values corresponding to the data points $\{d_i\}$. % please check to make sure intended meaning has been retained AUTHORS: Checked. It is ok.
 	The bottom panel shows the relative deviation between $d_{i}+ \epsilon _{b} \sigma _{bi}+ \epsilon _{c} \sigma _{c}d_{i}$, the $d\sigma/dt$ data points $d_i$ shifted within errors by the correlation parameters of the $\chi^2$ definition Equation~(\ref{eq:chi2-final}), and $m_i$, the $d\sigma/dt$ values calculated from the model.}}
	\label{fig:a}
\end{figure}

\begin{figure}[!hbt]
	\centering	\includegraphics[width=0.9\linewidth]{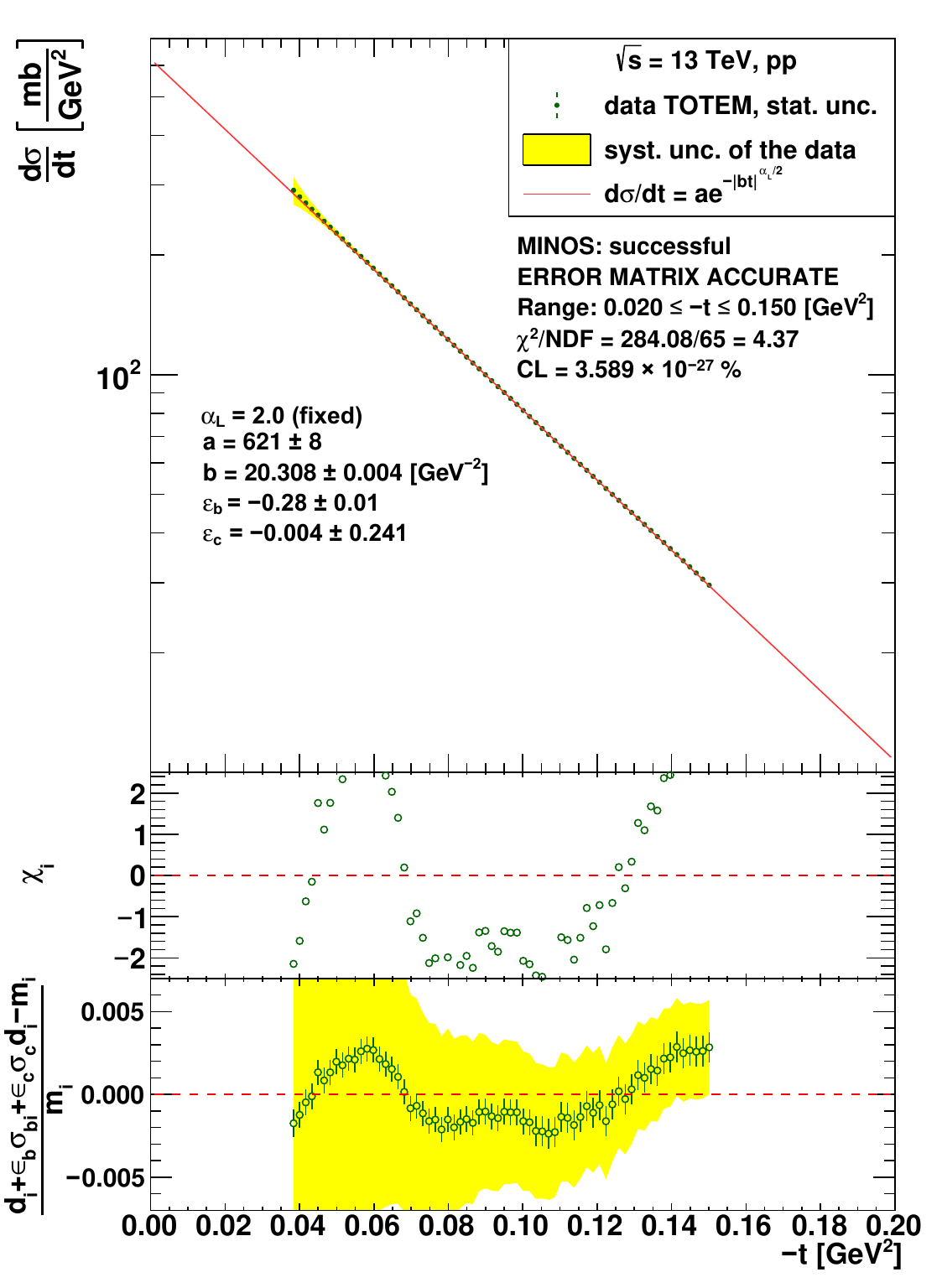}
	\caption{Same %MDPI: 1. Please change the hyphen (-) into a minus sign ($-$, "U+2212"). e.g., "-1" should be "$-$1". 2. Please change the terms into scientific notations in the figure,  e.g., "$8 \times 10^{3}$", not "8E3". AUTHORS: Done.
 as Figure~\ref{fig:1} but with $\alpha_L=2$ fixed.}
	\label{fig:b}
\end{figure}

As an illustrative example, the SL model fit to the most precise TOTEM data measured at $\sqrt s =$ 13 TeV \cite{TOTEM:2018hki} is shown in Figure~\ref{fig:a}, and the case with $\alpha_L=2$ fixed is shown in Figure~\ref{fig:b}. The SL model with $\alpha_L=1.958\pm0.003$ describes the  13 TeV TOTEM data with CL = 89.12\%, while the $\alpha_L=2$ fixed case fit has a confidence level of 3.6 $\times~10^{-27}$\%. These values are not surprising if one compares the bottom panel of Figure~\ref{fig:a} to the bottom panel of Figure~\ref{fig:b}. {This result clearly shows the success of the non-exponential SL model with $\alpha_L<2$.}

\vspace{-8pt}

Looking at the bottom panel of Figure~\ref{fig:a}, one can observe some oscillations in the data. This oscillation is a significant effect when only the statistical errors are considered. If systematic errors are taken into account {as well}, this oscillation effect {disappears}. {The SL model has a monotonically decreasing shape and describes the data with a good confidence level. This excludes the statistically significant oscillatory behavior of the data.}  

{Let us now discuss the energy dependence of the SL model parameters.} According to our analysis, surprisingly, the $\alpha_L$ parameter of the SL model is energy-independent, and {its value is 1.959 $\pm$ 0.002, indicating a L\'evy-$\alpha$-stable-shaped, power-law tail feature for impact parameter profiles $\tilde\sigma_{\rm in}$ and $\Gamma$ in the energy range 546 GeV $\leq \sqrt{s} \leq$ 13 TeV.}

We showed in Section~\ref{sec:fits} that the energy dependence of the optical point parameter of the SL model is compatible with a quadratically logarithmic shape; however, the $a$ parameter values determined from ATLAS and TOTEM data on $pp$ elastic scattering disagree. This discrepancy is a well-known fact, and the interpretation is that the ATLAS and TOTEM experiments use different methods to obtain the absolute normalization of the measurements \cite{ATLAS:2022mgx}.

%In Ref.~\cite{Csorgo:2023pdn}, we discussed that the optical point parameter is related to the $\alpha$ or opacity parameter of the LBB model which regulates the magnitude of the real size of the elastic scattering amplitude. Note that this opacity parameter $\alpha$ is not to be confused with the Levy index of stability $\alpha_L$, where the subscript $\null_L$ stands for  Lévy.   The value of the opacity parameter $\alpha$ is different in $pp$ and $p\bar p$ elastic scattering at the same energies. This implies different values for $pp$ and $p\bar p$ optical points too. Such a conclusion is seemingly in disagreement with the result of Section~\ref{sec:endep} that $p\bar p$ and $pp$ $a$ parameter values lie in the same curve. There is no real contradiction, only the precision of the measurements is too low to see the difference between $pp$ and $p\bar p$ optical points experimentally.

%According to our results presented in Ref.~\cite{Csorgo:2023pdn}, the slope parameter of the SL model can be written in terms of the parameters of the LBB model that have the same values in  $pp$ and $p\bar p$ elastic scattering at the same energies. This implies that slope parameters extracted from $pp$ and $p\bar p$ data should lie in the same energy dependence curve. Such a conclusion is again seemingly in disagreement with the result of Section~\ref{sec:endep}. 
We saw in Section~\ref{sec:fits} that $pp$ and $p\bar p$ $b$ parameter values lie in different curves. %There is no real contradiction again. 
There {are} two alternative interpretations.
The TOTEM Collaboration discussed in Ref.~\cite{TOTEM:2017asr} that there is a jump in the energy dependence of the slope parameter in the energy interval of 3~GeV $\lesssim \sqrt s \lesssim$ 4 GeV. One of the possibilities is  that the same jump is seen in our analysis {as well}, preventing the lower energy $p\bar p$ data to lie in the same curve with the higher energy LHC ATLAS and TOTEM data. The second possibility is that the difference between the $pp$ and $p\bar p$ slope is the effect of the odderon {\cite{CsT:ISMD}}. New $pp$ low-$|t|$ measurements at LHC at $\sqrt s =$ 1.8 and/or 1.96 TeV may clarify this issue.

%Finally, an important thing to note is that in Ref.~\cite{Csorgo:2023pdn}, applying approximations valid at low-$|t|$,
%we arrived to the conclusion that the total cross sections of $pp$ and $p\bar p$ scattering are the same. Such a result is in agreement with the strong form of the Pomeranchuk theorem stating that the difference of $pp$ and $p\bar p$ total cross sections goes to zero at asymptotic energies \cite{Barone:2002cv}. 

\section{Summary} \label{sec:summ}

We fitted the $pp$ and $p\bar p$ elastic differential cross section with a simple L\'evy $\alpha$-stable model in the center of mass energy range of 0.54~TeV $\leq\sqrt{s}\leq 13$ TeV and in the four-momentum transfer range of 0.02 GeV$^2$ $\leq -t\leq0.15$ GeV$^2$. We determined the energy dependence of the three parameters of the model. The L\'evy index of stability, $\alpha_L(s)$, results are consistent with an energy-independent, constant value that is slightly but significantly smaller than 2. The energy dependence of the optical point parameter is the same for $pp$ and $p\bar p$ processes and has a quadratically logarithmic shape; however, because of normalization differences, TOTEM and ATLAS optical point data {are inconsistent} within experimental errors. Thus, they can be fitted separately from one another and, furthermore, both can be fitted together with $p\bar p$ data; however, the ATLAS and TOTEM optical points cannot be fitted together, neither without nor with the $p\bar p$  data.

 In our L\'evy analysis, we observe that the Lévy slope parameter $b(s)$ has different energy dependence for $pp$ and $p\bar p$ scattering. This may be an odderon signal \cite{CsT:ISMD} or the ``jumping'' behavior in the energy interval of 3 GeV $\lesssim \sqrt s \lesssim$ 4 GeV as discussed by TOTEM in Ref.~\cite{TOTEM:2017asr}. New $pp$ low-$|t|$ measurements at LHC at $\sqrt s =$ 1.8 and/or 1.96 TeV may decide which interpretation is true.  We also find that TOTEM and ATLAS slope parameter data can be fitted together with a linearly logarithmic shape, indicating that TOTEM and ATLAS data {differ} only in their normalization, {while} their shape is consistent. Similar conclusions were drawn in Ref.~\cite{Petrov:2023mww} concerning the TOTEM--ATLAS discrepancy at 13 TeV.

\section*{Acknowledgements}

We gratefully acknowledge inspiring discussions with A. Ster and V. Petrov and the support from NKFIH Grants no. K133046, K147557, and 2020-2.2.1-ED-2021-00181; MATE KKP 2023; MATE KKP 2024; \'UNKP-23-3 New National Excellence Program of the Ministry for Culture and Innovation from the source of the National Research, Development and Innovation Fund.

\appendix
\section*{Appendix: Simple Models of Elastic Scattering} \label{app:simpmodels}
\renewcommand{\thesubsection}{\Alph{subsection}}

In this Appendix, we first recapitulate the basic quantities that characterize elastic scattering; then, we shortly review simple models of elastic scattering. 

The number one quantity that characterizes elastic scattering is the differential cross section, as given by Equation~(\ref{eq:diffxsec}). The elastic cross section is the integral of the elastic differential cross section over the Mandelstam variable $t$,
\begin{equation}\label{eq:elasticxsec}
  \sigma_{\rm el}(s) = \int_{-\infty}^0 {\rm d}t \frac{{\rm d}\sigma}{{\rm d} t}(s,t) .
\end{equation} %Authors: indentation is not needed. Valid for all below.
The optical point parameter, $a(s)$, is determined as the extrapolated value of the differential cross section to the optical point, $t = 0$,
\begin{equation}\label{eq:As}
  a(s) = \lim_{t\rightarrow  0} \frac{{\rm d}\sigma}{{\rm d} t}(s,t) .
\end{equation}
The elastic slope is, in general, an $s$- and $t$-dependent function, defined as the logarithmic derivative of the differential cross section,
\begin{equation}\label{eq:Bst}
  B(s,t)   =  \frac{\rm d}{{\rm d} t} \ln \frac{{\rm d}\sigma}{{\rm d} t}(s,t).
\end{equation}
The ratio of the real to the imaginary part of the scattering amplitude, $\rho(s,t)$, is, in general, also an $s$- and $t$-dependent function,
\begin{equation}\label{eq:rhost}
  \rho(s,t)   =  \frac{ {\rm Re}T_{\rm el}(s,t)}{{\rm Im}T_{\rm el}(s,t)}.
\end{equation}
Both the $(s,t)$-dependent elastic slope and the real-to-imaginary ratio are frequently characterized by their values extrapolated to the optical point
that become quantities that depend only on the Mandelstam variable $s$:
\begin{eqnarray}
  B_0(s) \equiv B(s) \, = \,  \lim_{t\rightarrow  0} B(s,t) ,
    \label{eq:Bs} \\
  \rho_0(s) \equiv \rho(s) \, = \,  \lim_{t\rightarrow  0} \rho(s,t) .
    \label{eq:rhos}
\end{eqnarray}
The optical theorem connects the total cross section, {i.e.,} the sum of the elastic and inelastic scattering cross sections, to the imaginary part of the elastic scattering amplitude at the vanishing four-momentum transfer:
\begin{equation}\label{eq:optical-theorem}
   \sigma_{\rm tot}(s) =  \sigma_{\rm el}(s)  + \sigma_{\rm in}(s) \, = \,  2 Im  T_{\rm el}(s,t=0) .
\end{equation}
This optical theorem follows from the unitarity of the scattering or $S$-matrix. Taken together with Equations~(\ref{eq:diffxsec}), (\ref{eq:As}), and (\ref{eq:rhos}), the optical theorem of Equation~(\ref{eq:optical-theorem}) implies a connection between the differential cross section at the vanishing four-momentum transfer, the total cross section, and the real-to-imaginary ratio as
\begin{equation}\label{eq:optical-theorem-relation}
    a(s) = \frac{1}{16 \pi } (1 +  \rho^2(s) )  \sigma^2_{\rm tot}(s).
\end{equation}

\subsection{Black Disc Model}

At asymptotically high center of mass energies, in the limit of $\sqrt{s} \rightarrow \infty$, many high-energy scattering models approach the so-called black disc limit.
As it is well known \cite{Block:2006hy,Barone:2002cv}, this choice corresponds to 
\begin{equation}
\Gamma(s, b) = \Theta\left(R(s) - b\right),
\end{equation}
where $R(s)$ stands for the radius of a black disc, $\Theta(x)=1$ for $x\geq0$, and $\Theta(x)=0$ for $x<0$. For a vanishing real part of the elastic scattering amplitude, Equations~(\ref{eq:impamp}) and (\ref{eq:relPWtoeik_2})
imply that 
\begin{equation}
    \Tilde t_{\rm el} (s,b)  = i \Theta\left(R(s) - b\right)     
\end{equation}
and
\begin{equation}
     T_{\rm el}(s,t) = 2\pi iR^2(s)\frac{J_1\left(qR(s)\right)}{qR(s)},
\end{equation}
where $J_1(x)$ is the Bessel function of the first kind and $-t = q^2$ relates the Mandelstam variable $t$ to the modulus of the transferred four-momentum, $q$.
Thus, the cross sections and the slope parameter are given as
\begin{eqnarray}
    \frac{{\rm d}\sigma}{{\rm d} t}(s,t) & = & \pi R^4(s) \left(\frac{J_1\left(q R(s)\right)}{q R(s)}\right)^2 \, , \\
    \sigma_{\rm el}(s) & = & \pi R^2(s)  \, , \\
    \sigma_{\rm in}(s) & = & \pi  R^2(s)  \, , \\
    \sigma_{\rm tot}(s) & = & 2  \pi R^2(s) \, , \\
     B_0(s)& = &\frac{R^2(s)}{4} \, .
\end{eqnarray}
In this black disc limit, half of the total cross section corresponds to elastic scattering:
\begin{equation}
    \frac{ \sigma_{\rm el}(s) }{ \sigma_{\rm tot}(s)} = \frac{1}{2} .
\end{equation}
The black disc model obeys a nearly exponential shape at small values of the four-momentum transfer $-t$. 

One can re-express the black disc model amplitudes and the differential cross section in terms of $\sigma_{\rm tot}$ and $B_0$
as follows:
\begin{equation}
    \tilde t_{\rm el}(s,b) = i\Theta\left(2\sqrt{B_0(s)}-b\right),
\end{equation}
\begin{equation}
     T_{\rm el}(s,t) = i\sigma_{\rm tot}(s)\frac{J_1\left(2q\sqrt{B_0(s)}\right)}{2q\sqrt{B_0(s)}},
\end{equation}
\begin{equation}
    \frac{{\rm d}\sigma}{{\rm d} t}(s,t)  =  \frac{1}{4\pi}  \sigma_{\rm tot}^2(s)\left(\frac{J_1\left(2q\sqrt{B_0(s)}\right)}{2q\sqrt{B_0(s)}}\right)^2,
\end{equation}
yielding
\begin{equation}
    a(s) = \frac{1}{16\pi}\sigma_{\rm tot}^2(s).
\end{equation}

\subsection{Gray Disc Model}

Although asymptotically expected, the elastic to total cross section ratio remains significantly smaller from the black disc value, even at the currently highest colliding energies of $\sqrt{s} = 13$ $-$ 13.6 TeV at the LHC. As detailed in Figure 6 of  Ref.~\cite{TOTEM:2017asr},
the elastic to total cross section ratio is an increasing function of $\sqrt{s}$ in the TeV energy range, and it crosses the important limit~\cite{Broniowski:2018xbg} of $ \frac{ \sigma_{\rm el}(s) }{ \sigma_{\rm tot}(s)} = \frac{1}{4}$ between 2.76 and 7 TeV. However, it is still about a factor of two smaller as compared to the black disc limit of $ \frac{ \sigma_{\rm el}(s) }{ \sigma_{\rm tot}(s)} = \frac{1}{2}$. The gray disc model improves on the black disc model by introducing a grayness parameter $G_r(s)$ as
\begin{equation}
     \Tilde t_{\rm el} (s,b) = 2 i G_r(s) \Theta\left(R(s) - b\right), 
\end{equation}
which yields 
\begin{eqnarray}
    T_{\rm el}(s,t) & = & 4\pi iR^2(s)G_r(s)\frac{J_1\left(qR(s)\right)}{qR(s)}. 
\end{eqnarray}
Then,
\begin{equation}
  \frac{{\rm d}\sigma}{{\rm d} t}(s,t)  =  4 \pi G^2_r(s) R^4(s)   \left(\frac{J_1\left(q R(s)\right)}{q R(s)}\right)^2 ,
\end{equation}
\begin{equation}
    \sigma_{\rm el}(s) = 4 \pi G^2_r(s) R^2(s) ,
\end{equation}
\begin{equation}
    \sigma_{\rm in}(s) =4 \pi G_r(s)\left(1 - G_r(s)\right) R^2(s),
\end{equation}
\begin{equation}
    \sigma_{\rm tot}(s)  =  4  \pi G_r(s) R^2(s), 
\end{equation}
\begin{equation}
    B_0(s) = \frac{R^2(s)}{4} = \frac{1}{16\pi}\frac{\sigma_{\rm tot}^2(s)}{\sigma_{\rm el}(s)},
\end{equation}
and
\begin{equation}\label{eq:tuneratio}
    G_r(s)=\frac{ \sigma_{\rm el}(s) }{ \sigma_{\rm tot}(s)} =  \frac{1}{16\pi}\frac{\sigma_{\rm tot}(s)}{B_0(s)}. 
\end{equation}
Given that the elastic cross section cannot be greater than the total cross section, it follows that the grayness parameter $G_r(s)$ is bound in the region of
$0 \leq G_r(s) \leq 1$ and the conventional black disc limit corresponds to the $G_r(s) = 1/2$ case.
With this grayness parameter, the elastic to total cross section ratio can be tuned to the measured experimental data. Two simple shortcomings of the model still remain. The first is the problem that the real part of the scattering amplitude vanishes; hence,  $\rho_0(s)= 0$, which is at variance with the experimental observations that indicate a small but non-vanishing real-to-imaginary ratio $\rho_0(s)$, even at the TeV energy range~\cite{TOTEM:2017sdy}. 

One can re-express the gray disc model amplitudes and the differential cross section in terms of $B_0$ and $\sigma_{\rm tot}$ as follows:
\begin{equation}
    \tilde t_{\rm el}(s,b) = \frac{i}{8\pi}\frac{\sigma_{\rm tot}(s)}{B_0(s)}\Theta\left(2\sqrt{B_0(s)}-b\right),
\end{equation}
\begin{equation}
     T_{\rm el}(s,t) = i\sigma_{\rm tot}(s)\frac{J_1\left(2q\sqrt{B_0(s)}\right)}{2q\sqrt{B_0(s)}},
\end{equation}
\begin{equation}
    \frac{{\rm d}\sigma}{{\rm d} t}(s,t)  =  \frac{1}{4\pi}  \sigma_{\rm tot}^2(s)\left(\frac{J_1\left(2q\sqrt{B_0(s)}\right)}{2q\sqrt{B_0(s)}}\right)^2.
\end{equation}

\subsection[\appendixname~\thesubsection]{Gray Disc Model with a Small Real Part}
The shortcoming of the gray disc model amplitude of having only an imaginary part can be easily fixed. A small, $s$-dependent real part can be trivially added to the gray disc model as follows:
\begin{equation}
     \Tilde t_{\rm el} (s,b) = 2(i+\rho_0(s)) G_r(s) \Theta\left(R(s) - b\right) ,
\end{equation}
which yields the following cross section relations:
\begin{eqnarray}
    \frac{{\rm d}\sigma}{{\rm d} t}(s,t) & = & 2 \pi( 1 + \rho^2_0(s) )   G^2_r(s) R^2(s) \left(\frac{J_1\left(q R(s)\right)}{q R(s)}\right)^2 \, , \\
    \sigma_{\rm el}(s) & = & 4 \pi( 1 + \rho^2_0(s) )  G^2_r(s) R^2(s)  \, ,  \\
    \sigma_{\rm in}(s) & = &  4 \pi G_r(s)\left(1 - (1 + \rho_0^2(s) ) \, G_r(s)\right) R^2(s)  \, ,  \\
    \sigma_{\rm tot}(s) & = & 4  \pi G_r(s) R^2(s) \, ,
\end{eqnarray}
which implies
\begin{equation}
   G_r(s)=\frac{1}{1+\rho^2_0(s)} \frac{ \sigma_{\rm el}(s) }{ \sigma_{\rm tot}(s)} . \label{eq:tuneratio-rho} 
\end{equation}
This is the generalization of Equation~(\ref{eq:tuneratio}) for the case of a gray disc with a non-vanishing real part. Thus, the ratio 
of elastic to total cross section of a gray disc with a non-vanishing real part can be  
tuned to the measured values with the help of a grayness parameter $G_r(s)$ if a small correction to the
real to imaginary ratio $\rho_0(s)$ is properly taken into account, as given by Equation~(\ref{eq:tuneratio-rho}) .

 A more difficult-to-fix shortcoming of the black and gray disc models is that
the differential cross section of the black disc and the gray disc model are both proportional to the Bessel function of the first kind, which has an infinite number of  zeros that correspond to an 
infinite number of minima of the differential cross section as a function of $-t$. However, experimentally, only a single diffractive minimum is observed in elastic proton--proton $(pp)$ scattering in the TeV range. This alone excludes the validity of both the black disc and the gray disc models.

One can re-express the gray disc model amplitudes and the differential cross section in terms of $B_0$, $\sigma_{\rm tot}$, and $\rho_0$ as follows:
\begin{equation}
    \tilde t_{\rm el}(s,b) = \frac{i+\rho_0(s)}{8\pi}\frac{\sigma_{\rm tot}(s)}{B_0(s)}\Theta\left(2\sqrt{B_0(s)}-b\right)
\end{equation}
\begin{equation}
     T_{\rm el}(s,t) = \left(i+\rho_0(s)\right)\sigma_{\rm tot}(s)\frac{J_1\left(2q\sqrt{B_0(s)}\right)}{2q\sqrt{B_0(s)}},
\end{equation}
\begin{equation}
    \frac{{\rm d}\sigma}{{\rm d} t}(s,t)  =  \frac{1+\rho_0^2(s)}{4\pi}  \sigma_{\rm tot}^2(s)\left(\frac{J_1\left(2q\sqrt{B_0(s)}\right)}{2q\sqrt{B_0(s)}}\right)^2,
\end{equation}
yielding
\begin{equation}
    a(s) = \frac{1+\rho^2_0(s)}{16\pi}\sigma_{\rm tot}^2(s).
\end{equation}

\subsection{Gaussian Model}

In the gray Gaussian model, the profile function is Gaussian, and hence, the amplitude in the impact parameter representation has the form    
\begin{eqnarray}\label{eq:profgaussapp}
\tilde t_{\rm el}(s, b)  =  4i G_r(s) 
\exp\left({-\frac{b^2}{2 R^2_G(s)}}\right),
\end{eqnarray}
where $G_r(s)$ is an $s$-dependent grayness factor as before and $R_G(s)$ is an  $s$-dependent Gaussian radius parameter.
A small and $t$-independent but $s$-dependent real part can also be added at this point, similarly to how this was performed in the case of the gray disc model.
In the small $-t$ approximation, corresponding to large values of the impact parameter  $b$, the gray Gaussian model assumption with an $s$-dependent real-to-imaginary ratio yields the following elastic scattering amplitudes:
\begin{eqnarray}
     \Tilde t_{\rm el} (s,b) & = & 4  (  i + \rho_0(s) ) \, G_r(s)
                        \exp{\left(-\frac{b^2}{2 R^2_G(s)}\right)} , \\
      T_{\rm el}(s,t)  & = & 8  \pi  (  i + \rho_0(s) ) \, G_r(s) R^2_G(s)
               \exp{  \left(\frac{t R^2_G(s)}{2}\right)}.
\end{eqnarray}
The root mean square of this distribution, corresponding to  $R_G(s)$, is usually identified with the proton's size, as resolved at a given energy. 

This so-called gray Gaussian model yields 
\begin{equation}\label{eq:SL0app}
\frac{{\rm d}\sigma}{{\rm d} t}(s,t) = 
   16  \pi \,  \left(1 + \rho_0^2(s) \right)   \,  G^2_r(s) \, R^4_G(s) \,               \exp{  \left(t R^2_G(s)\right)},
\end{equation}
\begin{equation}
    \sigma_{\rm el}(s) = 16\pi \left(1 + \rho_0^2(s) \right)  G^2_r(s) R_G^2(s) ,
\end{equation}
\begin{equation}
    \sigma_{\rm in}(s) =   16 \pi G_r(s)\left(1 - \left(1 + \rho_0^2(s) \right) \, G_r(s)\right) R_G^2(s) ,
\end{equation}
\begin{equation}
    \sigma_{\rm tot}(s)  =  16  \pi G_r(s) R_G^2(s), 
\end{equation}
\begin{equation}
    B_0(s) = R_G^2(s) .
\end{equation}
%Thus, for Gaussian sources and the corresponding exponential differential cross-sections at small $-t$, the (nuclear) slope parameter $B_0(s)$ is a measure of the root mean square of the inelastic scattering distribution in the impact parameter space.
The above formulae generalize the equivalent formulae obtained for a gray disc and are valid in the diffractive cone only. This way, the problem of infinitely many diffractive minima obtained for a sharply cut black or gray disc is eliminated. One can conclude that the gray Gaussian model has two remaining shortcomings: (i) it does not describe the single diffractive minimum seen in the data, and (ii) it gives a purely exponential low-$|t|$ differential cross section, contradicting the most precise measurements at LHC \cite{TOTEM:2015oop}.

In this case, the grayness factor can also be expressed as
\begin{equation}
 G_r(s) = \frac{1}{1+\rho^2_0(s)} \frac{ \sigma_{\rm el}(s) }{ \sigma_{\rm tot}(s)}  = 
    \frac{1}{16 \pi }  \,   \frac{\sigma_{\rm tot}(s) }{B_0(s)}  . \label{eq:tuneratio-rho-grey-gauss} 
\end{equation}

%According to the above equations, the optical point $A(s)$ measures a combination of the greyness factor $G_r(s)$, the real-to-imaginary ratio  $\rho_0(s)$ and the Gaussian radius parameter $R_G(s)$.

We can re-express the amplitudes and the differential cross section in terms of $B_0$, $\rho_0$, and $\sigma_{\rm tot}$ as follows:
\begin{equation}
    \tilde t_{\rm el}(s,b) = \frac{i+\rho_0(s)}{4\pi}\frac{\sigma_{\rm tot}(s)}{B_0(s)}\exp{\left(-\frac{b^2}{2 B_0(s)}\right)},
\end{equation}
\begin{equation}
     T_{\rm el}(s,t) = \frac{i+\rho_0(s)}{2} \sigma_{\rm tot}(s)\exp{\left(\frac{tB_0(s)}{2 }\right)},
\end{equation}
\begin{equation}
    \frac{{\rm d}\sigma}{{\rm d} t}(s,t)  =   \frac{1+\rho^2_0(s)}{16\pi}\sigma_{\rm tot}^2(s)\exp{\left(tB_0(s)\right)},
\end{equation}
yielding
\begin{equation}
    a(s) = \frac{1+\rho^2_0(s)}{16\pi}\sigma_{\rm tot}^2(s).
\end{equation}

\subsection{L\'evy $\alpha$-Stable Model}

The L\'evy $\alpha$-stable model describes a non-exponential low-$|t|$ differential cross section.
In the gray L\'evy $\alpha$-stable model with a small real part, the profile function is L\'evy $\alpha$-stable distributed, and hence, the amplitudes have the form:
\begin{eqnarray}
     \Tilde t_{\rm el} (s,b) & = & \frac{i+\rho_0(s)}{2}\sigma_{\rm tot}(s) \frac{1}{4\pi^2}\int d^2\vec \Delta \exp{\left(\vec \Delta\cdot\vec b\right)}
                        \exp{\left(-\frac{1}{2}\big|\Delta^2R_L^2(s)\big|^{\alpha_L(s)/2}\right)} , \\
      T_{\rm el}(s,t)  & = & \frac{i+\rho_0(s)}{2} \sigma_{\rm tot}(s) 
               \exp{\left(-\frac{1}{2}\big|tR_L^2(s)\big|^{\alpha_L(s)/2}\right)}.
\end{eqnarray}
This gray L\'evy $\alpha$-stable model yields 
\begin{equation}\label{eq:SL0app}
\frac{{\rm d}\sigma}{{\rm d} t}(s,t) = 
\frac{1+\rho^2_0(s)}{16 \pi} \sigma^2_{\rm tot}(s) 
               \exp{\left(-\big|tR_L^2(s)\big|^{\alpha_L(s)/2}\right)},
\end{equation}
\begin{equation}
    \sigma_{\rm el}(s) = \frac{1+\rho^2_0(s)}{8 \pi \, \alpha_L(s)}  \, \Gamma\left(\frac{2}{\alpha_L(s)} \right)  \frac{\sigma^2_{\rm tot}(s) }{R^2_L(s)},
    %\left(1 + \rho_0^2(s) \right) ,
\end{equation}
and the inelastic cross section is given as the difference between the total and the elastic cross sections,
\begin{equation}
    \sigma_{\rm in}(s) =    \sigma_{\rm tot}(s) \left(
    1 - \frac{1+\rho^2_0(s)}{8 \pi \, \alpha_L(s)}  \, \Gamma\left(\frac{2}{\alpha_L(s)} \right)  \frac{\sigma_{\rm tot}(s) }{R^2_L(s)}
    \right).
\end{equation}

The (nuclear) slope parameter  $B_0(s)$ is divergent for a L\'evy source, but a generalized slope parameter can be introduced
that takes into account the non-exponential nature of the small $-t$ shape of the differential cross section.
This definition identifies a generalized $b_L(s)$, with the L\'evy scale parameter:
\begin{equation}
    b_{\rm L}(s |\alpha_L) =    R^2_{L}(s), 
\end{equation}
which is, in general, a measure of the scale of the source even in the general $ 0 < \alpha_L(s) \leq 2 $ case.
In the special Gaussian case of $\alpha_L = 2$, the differential cross section becomes nearly exponential 
and $ b_{\rm L}(s|\alpha_L = 2) = B_0(s)$ becomes a measure of the root mean square of the inelastic scattering distribution in the impact parameter space.

Finally, for a simple L\'evy source model, the grayness factor is given by
\begin{equation}
    G_r(s) = \frac{1}{1+\rho^2_0(s)} \frac{ \sigma_{\rm el}(s) }{ \sigma_{\rm tot}(s)} \, = \, 
    \frac{1}{8 \pi \, \alpha_L(s)}  \, \Gamma\left(\frac{2}{\alpha_L(s)} \right)  \frac{\sigma_{\rm tot}(s) }{R^2_L(s)}
    . \label{eq:tuneratio-rho-Levy} 
\end{equation}
This formula generalizes Equation~(\ref{eq:tuneratio-rho-grey-gauss}) for L\'evy stable source distributions. The Gaussian result,   Equation~(\ref{eq:tuneratio-rho-grey-gauss}),
is recovered in the $\alpha_L = 2$ special case.

\bibliographystyle{ws-ijmpa}
\bibliography{Odderon-Letter.bib}

\providecommand{\noopsort}[1]{}\providecommand{\singleletter}[1]{#1}%
\begin{thebibliography}{10}
\expandafter\ifx\csname urlstyle\endcsname\relax
  \providecommand{\doi}[1]{doi:\discretionary{}{}{}#1}\else
  \providecommand{\doi}{doi:\discretionary{}{}{}\begingroup
  \urlstyle{rm}\Url}\fi

\bibitem{Barbiellini:1972ua}
G.~Barbiellini {\em et~al.}, {\em Phys. Lett. B} {\bf 39}, 663  (1972),
  \doi{10.1016/0370-2693(72)90025-1}.

\bibitem{Nagy:1978iw}
E.~Nagy {\em et~al.}, {\em Nucl. Phys. B} {\bf 150}, 221  (1979),
  \doi{10.1016/0550-3213(79)90301-8}.

\bibitem{Donnachie:1996rq}
A.~Donnachie and P.~V. Landshoff, {\em Phys. Lett. B} {\bf 387}, 637  (1996),
  \href{http://arxiv.org/abs/hep-ph/9607377}{{\ttfamily arXiv:hep-ph/9607377}},
  \doi{10.1016/0370-2693(96)01065-9}.

\bibitem{TOTEM:2018psk}
 TOTEM Collaboration (G.~Antchev {\em et~al.}), {\em Eur. Phys. J. C} {\bf 80},
   ~91  (2020), \href{http://arxiv.org/abs/1812.08610}{{\ttfamily
  arXiv:1812.08610 [hep-ex]}}, \doi{10.1140/epjc/s10052-020-7654-y}.

\bibitem{TOTEM:2013lle}
 TOTEM Collaboration (G.~Antchev {\em et~al.}), {\em EPL} {\bf 101},   21002
  (2013), \doi{10.1209/0295-5075/101/21002}.

\bibitem{TOTEM:2015oop}
 TOTEM Collaboration (G.~Antchev {\em et~al.}), {\em Nucl. Phys. B} {\bf 899},
  527  (2015), \href{http://arxiv.org/abs/1503.08111}{{\ttfamily
  arXiv:1503.08111 [hep-ex]}}, \doi{10.1016/j.nuclphysb.2015.08.010}.

\bibitem{TOTEM:2016lxj}
 TOTEM Collaboration (G.~Antchev {\em et~al.}), {\em Eur. Phys. J. C} {\bf 76},
    661  (2016), \href{http://arxiv.org/abs/1610.00603}{{\ttfamily
  arXiv:1610.00603 [nucl-ex]}}, \doi{10.1140/epjc/s10052-016-4399-8}.

\bibitem{TOTEM:2021imi}
 TOTEM Collaboration (G.~Antchev {\em et~al.}), {\em Eur. Phys. J. C} {\bf 82},
    263  (2022), \href{http://arxiv.org/abs/2111.11991}{{\ttfamily
  arXiv:2111.11991 [hep-ex]}}, \doi{10.1140/epjc/s10052-022-10065-x}.

\bibitem{TOTEM:2017sdy}
 TOTEM Collaboration (G.~Antchev {\em et~al.}), {\em Eur. Phys. J. C} {\bf 79},
    785  (2019), \href{http://arxiv.org/abs/1812.04732}{{\ttfamily
  arXiv:1812.04732 [hep-ex]}}, \doi{10.1140/epjc/s10052-019-7223-4}.

\bibitem{TOTEM:2018hki}
 TOTEM Collaboration (G.~Antchev {\em et~al.}), {\em Eur. Phys. J. C} {\bf 79},
    861  (2019), \href{http://arxiv.org/abs/1812.08283}{{\ttfamily
  arXiv:1812.08283 [hep-ex]}}, \doi{10.1140/epjc/s10052-019-7346-7}.

\bibitem{ATLAS:2014vxr}
 ATLAS Collaboration (G.~Aad {\em et~al.}), {\em Nucl. Phys. B} {\bf 889}, 486
  (2014), \href{http://arxiv.org/abs/1408.5778}{{\ttfamily arXiv:1408.5778
  [hep-ex]}}, \doi{10.1016/j.nuclphysb.2014.10.019}.

\bibitem{ATLAS:2016ikn}
 ATLAS Collaboration (M.~Aaboud {\em et~al.}), {\em Phys. Lett. B} {\bf 761},
  158  (2016), \href{http://arxiv.org/abs/1607.06605}{{\ttfamily
  arXiv:1607.06605 [hep-ex]}}, \doi{10.1016/j.physletb.2016.08.020}.

\bibitem{ATLAS:2022mgx}
 ATLAS Collaboration (G.~Aad {\em et~al.}), {\em Eur. Phys. J. C} {\bf 83},
  441  (2023), \href{http://arxiv.org/abs/2207.12246}{{\ttfamily
  arXiv:2207.12246 [hep-ex]}}, \doi{10.1140/epjc/s10052-023-11436-8}.

\bibitem{Breakstone:1985pe}
A.~Breakstone {\em et~al.}, {\em Phys. Rev. Lett.} {\bf 54},   2180  (1985),
  \doi{10.1103/PhysRevLett.54.2180}.

\bibitem{UA4:1985oqn}
 UA4 Collaboration (M.~Bozzo {\em et~al.}), {\em Phys. Lett. B} {\bf 155}, 197
  (1985), \doi{10.1016/0370-2693(85)90985-2}.

\bibitem{UA4:1986cgb}
 UA4 Collaboration (D.~Bernard {\em et~al.}), {\em Phys. Lett. B} {\bf 171},
  142  (1986), \doi{10.1016/0370-2693(86)91014-2}.

\bibitem{D0:2012erd}
 D0 Collaboration (V.~M. Abazov {\em et~al.}), {\em Phys. Rev. D} {\bf 86},
  012009  (2012), \href{http://arxiv.org/abs/1206.0687}{{\ttfamily
  arXiv:1206.0687 [hep-ex]}}, \doi{10.1103/PhysRevD.86.012009}.

\bibitem{UA4:1983mlb}
 UA4 Collaboration (R.~Battiston {\em et~al.}), {\em Phys. Lett. B} {\bf 127},
   472  (1983), \doi{10.1016/0370-2693(83)90296-4}.

\bibitem{UA4:1984skz}
 UA4 Collaboration (M.~Bozzo {\em et~al.}), {\em Phys. Lett. B} {\bf 147}, 392
  (1984), \doi{10.1016/0370-2693(84)90139-4}.

\bibitem{Cohen-Tannoudji:1972gqd}
G.~Cohen-Tannoudji, V.~V. Ilyin and L.~L. Jenkovszky, {\em Lett. Nuovo Cim.}
  {\bf 5S2}, 957  (1972), \doi{10.1007/BF02777999}.

\bibitem{Anselm:1972ir}
A.~A. Anselm and V.~N. Gribov, {\em Phys. Lett. B} {\bf 40}, 487  (1972),
  \doi{10.1016/0370-2693(72)90559-X}.

\bibitem{Tan:1974gd}
C.-I. Tan and D.~M. Tow, {\em Phys. Lett. B} {\bf 53}, 452  (1975),
  \doi{10.1016/0370-2693(75)90216-6}.

\bibitem{Khoze:2000wk}
V.~A. Khoze, A.~D. Martin and M.~G. Ryskin, {\em Eur. Phys. J. C} {\bf 18}, 167
   (2000), \href{http://arxiv.org/abs/hep-ph/0007359}{{\ttfamily
  arXiv:hep-ph/0007359}}, \doi{10.1007/s100520000494}.

\bibitem{Jenkovszky:2014yea}
L.~Jenkovszky and A.~Lengyel, {\em Acta Phys. Polon. B} {\bf 46}, 863  (2015),
  \href{http://arxiv.org/abs/1410.4106}{{\ttfamily arXiv:1410.4106 [hep-ph]}},
  \doi{10.5506/APhysPolB.46.863}.

\bibitem{Fagundes:2015vva}
D.~A. Fagundes, L.~Jenkovszky, E.~Q. Miranda, G.~Pancheri and P.~V. R.~G.
  Silva, { {Fine structure of the diffraction cone: from ISR to the LHC}}, in
  {\em {Gribov-85 Memorial Workshop on Theoretical Physics of XXI Century}\/},
  (9 2015).
\newblock \href{http://arxiv.org/abs/1509.02197}{{\ttfamily arXiv:1509.02197
  [hep-ph]}}.

\bibitem{Jenkovszky:2017efs}
L.~Jenkovszky, I.~Szanyi and C.-I. Tan, {\em Eur. Phys. J. A} {\bf 54},   116
  (2018), \href{http://arxiv.org/abs/1710.10594}{{\ttfamily arXiv:1710.10594
  [hep-ph]}}, \doi{10.1140/epja/i2018-12567-5}.

\bibitem{Kohara:2019qoq}
A.~K. Kohara, {\em J. Phys. G} {\bf 46},   125001  (2019),
  \href{http://arxiv.org/abs/1906.01402}{{\ttfamily arXiv:1906.01402
  [hep-ph]}}, \doi{10.1088/1361-6471/ab47d3}.

\bibitem{Kohara:2018wng}
A.~K. Kohara, E.~Ferreira and M.~Rangel, {\em Phys. Lett. B} {\bf 789}, 1
  (2019), \href{http://arxiv.org/abs/1811.03212}{{\ttfamily arXiv:1811.03212
  [hep-ph]}}, \doi{10.1016/j.physletb.2018.12.021}.

\bibitem{Csorgo:2023pdn}
T.~Cs\"org\H{o}, S.~Hegyi and I.~Szanyi, {\em Universe} {\bf 9},   361  (2023),
  \href{http://arxiv.org/abs/2308.05000}{{\ttfamily arXiv:2308.05000
  [hep-ph]}}, \doi{10.3390/universe9080361}.

\bibitem{uchaikin2011chance}
V.~V. Uchaikin and V.~M. Zolotarev, {\em Chance and stability: stable
  distributions and their applications} (Walter de Gruyter, 2011).

\bibitem{Tsallis:1995zz}
C.~Tsallis, S.~V.~F. Levy, A.~M.~C. Souza and R.~Maynard, {\em Phys. Rev.
  Lett.} {\bf 75}, 3589  (1995), \doi{10.1103/PhysRevLett.75.3589}.

\bibitem{Prato:1999jj}
D.~Prato and C.~Tsallis, {\em Phys. Rev. E} {\bf 60},   2398  (1999),
  \doi{10.1103/PhysRevE.60.2398}.

\bibitem{nolan2020univariate}
J.~P. Nolan, {\em Univariate stable distributions} (Springer, 2020).

\bibitem{Wilk:1999dr}
G.~Wilk and Z.~Wlodarczyk, {\em Phys. Rev. Lett.} {\bf 84},   2770  (2000),
  \href{http://arxiv.org/abs/hep-ph/9908459}{{\ttfamily arXiv:hep-ph/9908459}},
  \doi{10.1103/PhysRevLett.84.2770}.

\bibitem{Brax:1990jv}
P.~Brax and R.~B. Peschanski, {\em Phys. Lett. B} {\bf 253}, 225  (1991),
  \doi{10.1016/0370-2693(91)91388-C}.

\bibitem{Zhang:1995uf}
Y.~Zhang, L.-S. Liu and Y.-f. Wu, {\em Z. Phys. C} {\bf 71}, 499  (1996),
  \doi{10.1007/s002880050196}.

\bibitem{Csorgo:2003uv}
T.~Cs{\"o}rg{\H o}, S.~Hegyi and W.~A. Zajc, {\em Eur. Phys. J. C} {\bf 36}, 67
   (2004), \href{http://arxiv.org/abs/nucl-th/0310042}{{\ttfamily
  arXiv:nucl-th/0310042}}, \doi{10.1140/epjc/s2004-01870-9}.

\bibitem{Csorgo:2008ah}
T.~Cs{\"o}rg{\H o}, W.~Kittel, W.~J. Metzger and T.~Nov\'ak, {\em Phys. Lett.
  B} {\bf 663}, 214  (2008), \href{http://arxiv.org/abs/0803.3528}{{\ttfamily
  arXiv:0803.3528 [hep-ph]}}, \doi{10.1016/j.physletb.2008.04.029}.

\bibitem{L3:2011kzb}
 L3 Collaboration (P.~Achard {\em et~al.}), {\em Eur. Phys. J. C} {\bf 71},
  1648  (2011), \href{http://arxiv.org/abs/1105.4788}{{\ttfamily
  arXiv:1105.4788 [hep-ex]}}, \doi{10.1140/epjc/s10052-011-1648-8}.

\bibitem{Kurgyis:2020vbz}
B.~Kurgyis, D.~Kincses, M.~Nagy and M.~Csan\'ad, {\em Universe} {\bf 9},   328
  (2023), \href{http://arxiv.org/abs/2007.10173}{{\ttfamily arXiv:2007.10173
  [nucl-th]}}, \doi{10.3390/universe9070328}.

\bibitem{PHENIX:2017ino}
 PHENIX Collaboration (A.~Adare {\em et~al.}), {\em Phys. Rev. C} {\bf 97},
  064911  (2018), \href{http://arxiv.org/abs/1709.05649}{{\ttfamily
  arXiv:1709.05649 [nucl-ex]}}, \doi{10.1103/PhysRevC.97.064911}, [Erratum:
  Phys.Rev.C 108, 049905 (2023)].

\bibitem{Schegelsky:2018tit}
V.~A. Schegelsky, {\em Phys. Part. Nucl. Lett.} {\bf 16}, 503  (2019),
  \href{http://arxiv.org/abs/1804.07153}{{\ttfamily arXiv:1804.07153
  [hep-ph]}}, \doi{10.1134/S1547477119050261}.

\bibitem{CMS:2023xyd}
 CMS Collaboration (A.~Tumasyan {\em et~al.}) (6 2023),
  \href{http://arxiv.org/abs/2306.11574}{{\ttfamily arXiv:2306.11574
  [nucl-ex]}}.

\bibitem{Lokos:2018dqq}
 PHENIX Collaboration (S.~L\"ok\"os), {\em Universe} {\bf 4},  ~31  (2018),
  \href{http://arxiv.org/abs/1801.08827}{{\ttfamily arXiv:1801.08827
  [nucl-ex]}}, \doi{10.3390/universe4020031}.

\bibitem{Kincses:2024sin}
STAR Collaboration, D.~Kincses, { {Pion interferometry with L\'evy-stable
  sources in $\sqrt{s_{NN}}$ = 200 GeV Au+Au collisions at STAR}}, in {\em
  {52nd International Symposium on Multiparticle Dynamics}\/},  (1 2024).
\newblock \href{http://arxiv.org/abs/2401.11169}{{\ttfamily arXiv:2401.11169
  [nucl-ex]}}.

\bibitem{Porfy:2019scf}
 NA61/SHINE Collaboration (B.~P\'orfy), {\em Universe} {\bf 5},   154  (2019),
  \href{http://arxiv.org/abs/1906.06065}{{\ttfamily arXiv:1906.06065
  [nucl-ex]}}, \doi{10.3390/universe5060154}.

\bibitem{Porfy:2023yii}
 NA61/SHINE Collaboration (B.~Porfy), {\em Universe} {\bf 9},   298  (2023),
  \href{http://arxiv.org/abs/2306.08696}{{\ttfamily arXiv:2306.08696
  [nucl-ex]}}, \doi{10.3390/universe9070298}.

\bibitem{Csanad:2024hva}
M.~Csan\'ad and D.~Kincses, { {Femtoscopy with L\'evy sources from SPS through
  RHIC to LHC}}, in {\em {52nd International Symposium on Multiparticle
  Dynamics}\/},  (1 2024).
\newblock \href{http://arxiv.org/abs/2401.01249}{{\ttfamily arXiv:2401.01249
  [hep-ph]}}.

\bibitem{Novak:2016cyc}
T.~Nov{\'a}k, T.~Cs{\" o}rg{\H o}, H.~C. Eggers and M.~de~Kock, {\em Acta Phys.
  Polon. Supp.} {\bf 9},   289  (2016),
  \href{http://arxiv.org/abs/1604.05513}{{\ttfamily arXiv:1604.05513
  [physics.data-an]}}, \doi{10.5506/APhysPolBSupp.9.289}.

\bibitem{Csorgo:2019egs}
T.~Csörgő, R.~Pasechnik and A.~Ster, {\em Eur. Phys. J.} {\bf C80},   126
  (2020), \href{http://arxiv.org/abs/1910.08817}{{\ttfamily arXiv:1910.08817
  [hep-ph]}}, \doi{10.1140/epjc/s10052-020-7681-8}.

\bibitem{Csorgo:2018uyp}
T.~Cs{\"o}rg{\H o}, R.~Pasechnik and A.~Ster, {\em Eur. Phys. J.} {\bf C79},
  ~62  (2019), \href{http://arxiv.org/abs/1807.02897}{{\ttfamily
  arXiv:1807.02897 [hep-ph]}}, \doi{10.1140/epjc/s10052-019-6588-8}.

\bibitem{glauber1959lectures}
R.~Glauber, {\em Interscience, New York} {\bf 1},   315  (1959).

\bibitem{glauber1970theory}
R.~Glauber, { Theory of high energy hadron-nucleus collisions}, in {\em
  High-Energy Physics and Nuclear Structure: Proceedings of the Third
  International Conference on High Energy Physics and Nuclear Structure
  sponsored by the International Union of Pure and Applied Physics, held at
  Columbia University, New York City, September 8--12, 1969\/},  pp. 207--264.

\bibitem{Barone:2002cv}
V.~Barone and E.~Predazzi, {\em {High-Energy Particle Diffraction}} (Springer,
  Berlin, 2002).

\bibitem{Glauber:1984su}
R.~J. Glauber and J.~Velasco, {\em Phys. Lett. B} {\bf 147}, 380  (1984),
  \doi{10.1016/0370-2693(84)90137-0}.

\bibitem{Csorgo:2020wmw}
T.~Cs{\"o}rg{\H o} and I.~Szanyi, {\em Eur. Phys. J. C} {\bf 81},   611
  (2021), \href{http://arxiv.org/abs/2005.14319}{{\ttfamily arXiv:2005.14319
  [hep-ph]}}, \doi{10.1140/epjc/s10052-021-09381-5}.

\bibitem{Bialas:2006qf}
A.~Bialas and A.~Bzdak, {\em Acta Phys. Polon.} {\bf B38}, 159  (2007),
  \href{http://arxiv.org/abs/hep-ph/0612038}{{\ttfamily arXiv:hep-ph/0612038
  [hep-ph]}}.

\bibitem{Nemes:2015iia}
F.~Nemes, T.~Cs{\" o}rg{\H o} and M.~Csanád, {\em Int. J. Mod. Phys.} {\bf
  A30},   1550076  (2015), \href{http://arxiv.org/abs/1505.01415}{{\ttfamily
  arXiv:1505.01415 [hep-ph]}}, \doi{10.1142/S0217751X15500761}.

\bibitem{Szanyi:2022ezh}
I.~Szanyi and T.~Cs\"org\H{o}, {\em Eur. Phys. J. C} {\bf 82},   827  (2022),
  \href{http://arxiv.org/abs/2204.10094}{{\ttfamily arXiv:2204.10094
  [hep-ph]}}, \doi{10.1140/epjc/s10052-022-10770-7}.

\bibitem{Adare:2008cg}
 PHENIX Collaboration (A.~Adare {\em et~al.}), {\em Phys. Rev.} {\bf C77},
  064907  (2008), \href{http://arxiv.org/abs/0801.1665}{{\ttfamily
  arXiv:0801.1665 [nucl-ex]}}, \doi{10.1103/PhysRevC.77.064907}.

\bibitem{E-710:1990vqb}
 E-710 Collaboration (N.~A. Amos {\em et~al.}), {\em Phys. Lett. B} {\bf 247},
  127  (1990), \doi{10.1016/0370-2693(90)91060-O}.

\bibitem{Cheng:1969eh}
H.~Cheng and T.~T. Wu, {\em Phys. Rev. Lett.} {\bf 22},   666  (1969),
  \doi{10.1103/PhysRevLett.22.666}.

\bibitem{Cheng:1969bf}
H.~Cheng and T.~T. Wu, {\em Phys. Rev.} {\bf 182}, 1852  (1969),
  \doi{10.1103/PhysRev.182.1852}.

\bibitem{Cheng:1969ac}
H.~Cheng and T.~T. Wu, {\em Phys. Rev.} {\bf 182}, 1868  (1969),
  \doi{10.1103/PhysRev.182.1868}.

\bibitem{Cheng:1987ga}
H.~Cheng and T.~T. Wu, {\em {EXPANDING PROTONS: SCATTERING AT HIGH-ENERGIES}}
  1987.

\bibitem{Bourrely:1978da}
C.~Bourrely, J.~Soffer and T.~T. Wu, {\em Phys. Rev. D} {\bf 19},   3249
  (1979), \doi{10.1103/PhysRevD.19.3249}.

\bibitem{Bourrely:2014efa}
C.~Bourrely, J.~Soffer and T.~T. Wu, {\em Int. J. Mod. Phys. A} {\bf 30},
  1542006  (2015), \href{http://arxiv.org/abs/1405.6698}{{\ttfamily
  arXiv:1405.6698 [hep-ph]}}, \doi{10.1142/S0217751X15420063}.

\bibitem{Giordano:2012mn}
M.~Giordano, E.~Meggiolaro and N.~Moretti, {\em JHEP} {\bf 09},   031  (2012),
  \href{http://arxiv.org/abs/1203.0961}{{\ttfamily arXiv:1203.0961 [hep-ph]}},
  \doi{10.1007/JHEP09(2012)031}.

\bibitem{Giordano:2013iga}
M.~Giordano and E.~Meggiolaro, {\em JHEP} {\bf 03},   002  (2014),
  \href{http://arxiv.org/abs/1311.3133}{{\ttfamily arXiv:1311.3133 [hep-ph]}},
  \doi{10.1007/JHEP03(2014)002}.

\bibitem{donnachie_dosch_landshoff_nachtmann_2002}
S.~Donnachie, G.~Dosch, P.~Landshoff and O.~Nachtmann, {\em Pomeron Physics and
  QCD} (Cambridge University Press, 2002).

\bibitem{Patrignani_2016}
C.~Patrignani, {\em Chinese Physics C} {\bf 40},   100001 (Oct 2016),
  \doi{10.1088/1674-1137/40/10/100001}.

\bibitem{Martynov:2007dy}
E.~Martynov, {\em Phys. Rev. D} {\bf 76},   074030  (2007),
  \href{http://arxiv.org/abs/hep-ph/0703248}{{\ttfamily arXiv:hep-ph/0703248}},
  \doi{10.1103/PhysRevD.76.074030}.

\bibitem{Martynov:2007kn}
E.~Martynov and B.~Nicolescu, {\em Eur. Phys. J. C} {\bf 56}, 57  (2008),
  \href{http://arxiv.org/abs/0712.1685}{{\ttfamily arXiv:0712.1685 [hep-ph]}},
  \doi{10.1140/epjc/s10052-008-0629-z}.

\bibitem{Cudell:2005sg}
J.~R. Cudell, A.~Lengyel and E.~Martynov, {\em Phys. Rev. D} {\bf 73},   034008
   (2006), \href{http://arxiv.org/abs/hep-ph/0511073}{{\ttfamily
  arXiv:hep-ph/0511073}}, \doi{10.1103/PhysRevD.73.034008}.

\bibitem{Cahn:1982nr}
R.~Cahn, {\em Z. Phys. C} {\bf 15},   253  (1982), \doi{10.1007/BF01475009}.

\bibitem{Martynov:2013ana}
E.~Martynov, {\em Phys. Rev. D} {\bf 87},   114018  (2013),
  \href{http://arxiv.org/abs/1305.3093}{{\ttfamily arXiv:1305.3093 [hep-ph]}},
  \doi{10.1103/PhysRevD.87.114018}.

\bibitem{TOTEM:2017asr}
 TOTEM Collaboration (G.~Antchev {\em et~al.}), {\em Eur. Phys. J. C} {\bf 79},
    103  (2019), \href{http://arxiv.org/abs/1712.06153}{{\ttfamily
  arXiv:1712.06153 [hep-ex]}}, \doi{10.1140/epjc/s10052-019-6567-0}.

\bibitem{CsT:ISMD}
T.~Cs\"org\H{o} and I.~Szanyi, {\em {Cross-checking Odderon signals at small
  values of four-momentum transfer.}}
\newblock {Presentation at ISMD 2023, Gy\"ongy\"os, Hungary
  \url{https://indico.cern.ch/event/1258038/contributions/5537131/attachments/2702376/4690629/2023-08-24-Csorgo-ISMD-final.pdf}}.

\bibitem{Petrov:2023mww}
V.~A. Petrov and N.~P. Tkachenko (3 2023),
  \href{http://arxiv.org/abs/2303.01058}{{\ttfamily arXiv:2303.01058
  [hep-ph]}}.

\bibitem{Block:2006hy}
M.~M. Block, {\em Phys. Rept.} {\bf 436}, 71  (2006),
  \href{http://arxiv.org/abs/hep-ph/0606215}{{\ttfamily arXiv:hep-ph/0606215
  [hep-ph]}}, \doi{10.1016/j.physrep.2006.06.003}.

\bibitem{Broniowski:2018xbg}
W.~Broniowski, L.~Jenkovszky, E.~Ruiz~Arriola and I.~Szanyi, {\em Phys. Rev.}
  {\bf D98},   074012  (2018),
  \href{http://arxiv.org/abs/1806.04756}{{\ttfamily arXiv:1806.04756
  [hep-ph]}}, \doi{10.1103/PhysRevD.98.074012}.

\end{thebibliography}

\end{document}